\documentclass[10pt,draft]{article}
\usepackage{hyperref}
\usepackage{amsfonts}
\usepackage{amssymb}
\usepackage{amsthm}
\usepackage{amsmath} 
\usepackage{graphicx}
\usepackage{float}
\usepackage[labelfont=bf]{caption}
\usepackage{amsfonts}
\usepackage{amssymb}
\usepackage{amsthm}
\usepackage{amsmath} 
\usepackage{graphicx}

\usepackage{color}
\date{}
\newcommand{\sss}{\setcounter{equation}{0}}
\newtheorem{theorem}{THEOREM}[section]
\renewcommand{\thetheorem}{\arabic{section}.\arabic{theorem}}

\newtheorem{remark}[theorem]{REMARK}

\newtheorem{prop}[theorem]{PROPOSITION}

\renewcommand{\theequation}{\arabic{section}.\arabic{equation}}

\def\beq{\begin{equation}}
\def\ene{\end{equation}}
\def \ds{\displaystyle}
\newcommand{\bull}{\hfill $\Box$}
\begin{document}
\baselineskip=10 pt
\parskip= 6 pt
\title{The  $L^{p}$ boundedness of the wave operators  for matrix Schr\"{o}dinger equations
\thanks{ Research partially supported by projects PAPIIT-DGAPA UNAM IN103918 and IN 100321,
and SEP-CONACYT CB 2015, 254062.}}
\author{ Ricardo Weder\thanks{Fellow, Sistema
Nacional de Investigadores. Electronic mail: weder@unam.mx. Home page: https://www.iimas.unam.mx/rweder/rweder.html }\\Departamento de F\'{\i}sica Matem\'{a}tica,\\Instituto de Investigaciones en Matem\'{a}ticas Aplicadas y en Sistemas.\\Universidad Nacional Aut\'{o}noma de M\'{e}xico,\\Apartado Postal 20-126, Ciudad de M\'{e}xico, 01000, M\'{e}xico.}
\date{}
\maketitle

\begin{abstract}
We prove that the wave operators   for $n \times n$   matrix
Schr\"odinger equations on the half line, with general selfadjoint boundary condition, are bounded in the spaces $L^p(\mathbb R^+, \mathbb C^n), 1 < p < \infty, $  for
slowly decaying selfadjoint matrix potentials, $V, $ that satisfy $\int_{0}^{\infty
}\, (1+x) |V(x)|\, dx < \infty.$ Moreover, assuming that $\int_{0}^{\infty }\, (1+x^\gamma) |V(x)|\, dx < \infty, \gamma > \frac{5}{2},$ and that the scattering matrix is the identity at zero and infinite energy, we prove that the wave operators are bounded in $L^1(\mathbb R^+, \mathbb C^n),$ and in $L^\infty(\mathbb R^+, \mathbb C^n).$
 We also prove that the wave operators for  $n\times n$  matrix Schr\"odinger equations on the line are bounded in the   spaces $L^p(\mathbb R, \mathbb C^n), 1 < p < \infty, $ assuming that the perturbation consists of a point interaction at the origin  and of a potential, $\mathcal V,$ that satisfies the condition  $\int_{-^{\infty}}^{\infty}\,
(1+|x|)\, |\mathcal V(x)|\, dx < \infty.$  Further,  assuming that $\int_{-\infty}^{\infty }\, (1+|x|^\gamma) |\mathcal V(x)|\,dx < \infty, \gamma > \frac{5}{2},$ and that the scattering matrix is the identity at zero and infinite energy, we prove that the wave operators are bounded in $L^1(\mathbb R, \mathbb C^n),$ and in $L^\infty(\mathbb R, \mathbb C^n).$ We obtain our results for  $n\times n$ matrix Schr\"odinger equations on the line from the  results for  $2n\times 2n$ matrix Schr\"odinger equations on the half line.


\end{abstract}

\baselineskip=15pt


\section{Introduction.}

In this paper we consider the wave operators for the matrix Schr\"{o}dinger equation on the half line
with general selfadjoint boundary condition,
\begin{equation}
\begin{array}{c}\label{0.0}
i\ds  \frac{\partial}{\partial t}u\left(  t,x\right)  =\left(  -\frac{\partial^2}{\partial x^2}+V\left(
x\right)  \right)  u\left(  t,x\right)  ,\ t\in\mathbb{R},\,x\in\mathbb{R}%
^{+},\\
u\left(  0,x\right)  =u_{0}\left(  x\right)  ,\text{ }x\in\mathbb{R}^{+},
\end{array}
\ene
\beq
-B^{\dagger}u\left(  t,0\right)  +A^{\dagger}  \frac{d}{dx}u(t,0) =0.
 \label{MS}
\end{equation}
Here $\mathbb{R}^{+}:=(0,+\infty),$ $u(t,x)$ is a function from
$\mathbb{R}\times\mathbb{R}^{+}$ into $\mathbb{C}^{n},A,B$ are constant
$n\times n$ matrices, the potential $V$ is a $n\times n$ selfadjoint
matrix-valued function of $x$, %
\begin{equation}
V\left(  x\right)  =V^{\dagger}\left(  x\right)  ,\text{ }x\in\mathbb{R}^{+}.
\label{PotentialHermitian}%
\end{equation}
The dagger designates the matrix adjoint. Let us denote by $M_n$ the set of all $n\times n$ matrices. We  assume that $V$ is in the
Faddeev class $L_{1}^{1}(\mathbb R^+, M_n),$ i.e. that it is a Lebesgue measurable $n \times n $ matrix-valued
function and,

\begin{equation}
\int_{\mathbb R^+}\,(1+x)\,|V(x)|\,dx<\infty, \label{PotentialL11}%
\end{equation}
where by $|V|$ we denote the matrix norm of $V.$ The more general selfadjoint
boundary condition at $x=0$  has been extensively studied. It can  can be written in many  equivalent ways. See,
\cite{[5]} and \cite{WederBook}, \cite{Harmer},
 \cite{Kostrykin1}, and \cite{Kostrykin2}. For other formulations of the general selfadjoint boundary condition see \cite{rf}. In this paper we use the parametrization of the boundary condition given in  
 \cite{[5]}, and Section 3.4 of Chapter 3 of \cite{WederBook}.  We write  the boundary condition  as
in (\ref{MS}), with constant  matrices $A$ and $B$ satisfying,
\begin{equation}
B^{\dagger}A =A^{\dagger}B, \label{wcon1}%
\end{equation}
and%
\begin{equation}
A^{\dagger}A+B^{\dagger}B>0. \label{wcon2}%
\end{equation}
We prove that the wave operators for the $n \times n$ matrix Schr\"odinger equation  on the half line \eqref{0.0} with the general selfadjoint boundary condition \eqref{MS}, \eqref{wcon1}, and \eqref{wcon2}, are bounded in the spaces $L^p(\mathbb R^+, \mathbb C^n), 1 < p < \infty.$ For this purpose, we suppose  that the potential satisfies  \eqref{PotentialHermitian} and \eqref{PotentialL11}. Moreover, assuming that $\int_{0}^{\infty }\, (1+x^\gamma) |V(x)|\, dx < \infty, \gamma > \frac{5}{2},$ and that the scattering matrix is the identity at zero and infinite energy, we prove that the wave operators are bounded in $L^1(\mathbb R^+, \mathbb C^n),$ and in $L^\infty(\mathbb R^+, \mathbb C^n).$
 We also  prove that the wave operators for the $n\times n$ Schr\"odinger equation on the line, with a point interaction at the origin and a potential,   are bounded in  $L^p(\mathbb R, \mathbb C^n), 1 < p < \infty.$ We  assume  that the potential, that we denote by $\mathcal V,$ is selfadjoint, i.e., 
 $\mathcal V(x)= \mathcal V(x)^\dagger,$ and
 \beq\label{ddddd}
 \int_{\mathbb R}\,(1+|x|)\,|\mathcal V(x)|\,dx<\infty.
 \ene
 Further,  assuming that $\int_{-\infty}^{\infty }\, (1+|x|^\gamma) |\mathcal V(x)|\, dx < \infty, \gamma > \frac{5}{2},$ and that the scattering matrix is the identity at zero and infinite energy, we prove that the wave operators are bounded in $L^1(\mathbb R, \mathbb C^n),$ and in $L^\infty(\mathbb R, \mathbb C^n).$ 
 We obtain the boundedness of the wave operators on the line for a $n
\times n$ matrix Schr\"odinger equation  from  the boundedness of the wave operators   for
a $2n\times2n$ matrix Schr\"odinger  equation on the half line.

In the scalar case there are several results on  the boundedness of the wave operators on the line.
Recall that in the scalar case the potential is {\it generic} if the zero energy Jost solutions from the left and from the right are linearly independent, and that it is {\it exceptional} if the zero energy Jost solutions from the left and from the right are linearly dependent. In the exceptional case the stationary Schr\"odinger equation on the line \eqref{seline} with zero energy, $k^2=0,$  has a bounded solution, that is called a zero-energy resonance, or a half-bound state.
In   \cite{we-99} it was proven that the wave operators are bounded  in $L^p(\mathbb R), 1 < p < \infty,$ under the assumption,
\beq\label{pscalar}
\int_{\mathbb R}\,  (1+|x|)^\gamma |\mathcal V(x)|\ dx < \infty,
\ene
with $\gamma > 3/2$ in the  generic  case and $\gamma > 5/2 $ in the  exceptional case. Furthermore, in \cite{we-99} it was proven that in the exceptional case if the Jost solution from the left at zero energy tends to one as $x \to - \infty,$ then, the wave operators are 
bounded in $L^1(\mathbb R)$ and in $L^\infty(\mathbb R).$  
The paper  \cite{we-99}  used a constructive proof that allowed to obtain a detailed low-energy  expansion, but that  was  somehow more demanding concerning the decay of the potential.  In \cite{gy} the boundedness of the wave operators in  $L^p(\mathbb R), 1 < p < \infty,$ was proven assuming that \eqref{pscalar} holds with $\gamma=3$ in the generic case and $\gamma=4$ in the exceptional case, and that moreover, $\frac{d}{dx} \mathcal V(x)$ satisfies \eqref{pscalar} with $\gamma=2,$ both in the generic and the exceptional cases.  The boundedness of the wave operators in $L^p(\mathbb R), 1 < p < \infty,$ was proven in \cite{daf} assuming \eqref{pscalar} with $\gamma=1,$ in the generic case and with $\gamma=2$ in the exceptional case. Furthermore, in \cite{dmw} the boundedness of the wave operators in $L^p(\mathbb R), 1 < p < \infty,$
 was proven for a potential that is the sum of a regular potential that satisfies \eqref{pscalar} with $\gamma > 3/2$ and of a singular potential that is a sum of Dirac delta functions.  In \cite{cu} the boundedness of the wave operators was proven for the discrete Schr\"odinger equation on the line.
There is a very extensive literature on the $L^p-$ boundedness of the wave operators and on  the related problem of dispersive estimates. For surveys see
\cite{Fanelli} and \cite{Schlag}, and \cite{ya2020} for recent results. In these papers   also the results in the multidimensional case are discussed.  

The matrix Schr\"odinger  equations find their origin at the very beginning of quantum mechanics. They are important in the description of particles with internal structure like spin and isospin, in atoms, molecules and in nuclear physics, and also in the study systems of particles. A well known example is the Pauli equation, that is the equation for half-spin particles. For further applications and references see  \cite{ag.1}-\cite{AgrMarch}, \cite{cs}, \cite{fa.1}, \cite{mk.1}, \cite{ll} and \cite{ne}.

 Since a number of years there is a renew of  the interest in matrix Schr\"odinger equations 
due to the importance of these   equations for quantum
graphs. For example, see \cite{Berkolaio}-\cite{Boman},
\cite{Gutkin}, \cite{Kostrykin1,Kostrykin2}, and  \cite{Kuchment}-\cite{Kurasov2}, as well as  the references  quoted there. The matrix Schr\"{o}dinger equation with a diagonal potential corresponds to a star graph. Such a quantum graph describes the dynamics  of $n$ connected very thin quantum wires that form a
star-graph, that is, a graph with only one vertex and a finite number of edges
of infinite length. This situation appears, for example, in the design of
elementary gates in quantum computing, in quantum wires, and in nanotubes for microscopic electronic
devices. In these cases  strings of atoms can  form a star-shaped graph.
The analysis  of the most general boundary condition at the vertex  is  important in the applications to problems in physics.
 A relevant example is the Kirchoff boundary condition.
A quantum graph is an idealization of wires with a small cross section that
meet at vertices. The graph  is obtained in  the limit when the cross section of the wires goes to
zero. As it turns out, the boundary conditions on the graph's vertices depends on how the limit
is taken. A priori, all the boundary conditions in (\ref{MS}) can appear
in this limit procedure. See Section 7.5 of \cite{Berkolaio} for a detailed discussion of the extensive literature on this problem. Hence, it is relevant  to  study  the more general
selfadjoint boundary condition.

The boundedness of the wave operators in $L^p$ spaces is an important problem on itself, and it has important applications.  Let us elaborate on this point.   For any selfadjoint operator, $H,$ in a Hilbert space we denote by $\mathcal H_{\textrm{ac}}(H)$  the subspace of absolute continuity of $H$ and by $P_{\textrm{ac}}(H)$ the orthogonal projector onto $\mathcal H_{\textrm{ac}}(H).$  Moreover, for any pair $H,H_0$ of selfadjoint operators in a Hilbert space  the wave operators  are defined as follows,
$$
W_\pm(H,H_0):= \text{\rm s-}\lim_{t \to \pm \infty}\, e^{it H}\, e^{-it H_0}\, P_{\rm ac}(H_0),
$$
provided that the strong limits exist.The operator $H$ is the perturbed Hamiltonian, and the operator  $H_0$  is the unperturbed Hamiltonian. The wave operators $W_\pm(H,H_0)$ are said to be complete if their range is equal to $\mathcal H_{\rm ac}(H).$ In the theory of scattering, the scattering solutions to the interacting Schr\"odinger equation
\beq\label{aschr}
i\frac{\partial}{\partial t} u(t)= H u(t), \qquad u(0)= \varphi,
\ene
are defined as $ e^{-it H}\varphi,$ with $\varphi \in \mathcal H_{\rm ac}(H).$ It is a purpose of scattering theory to compare the behavior for large times 
of the scattering solutions $ e^{-it H}\varphi$ with  the scattering solutions for   free Schr\"odinger equation
\beq\label{fseq}
i\frac{\partial}{\partial t} v(t)= H_0 v(t), \qquad v(0)= \psi,
\ene
with Hamiltonian, $H_0,$ that are given by $ e^{-it H_0}\psi,$ with  $\psi \in \mathcal H_{\rm ac}(H_0).$ If the wave operators exist and are complete, all the scattering solutions $e^{-it H}\varphi,$ to the interacting Schr\"odinger equation behave for large positive and negative times as scattering solutions for the free Sch\"odinger equation,  
$$
\lim_{t \to \pm \infty} \| e^{-it H} \varphi - e^{-it H_0 } W_\pm(H,H_0)^\ast \varphi\|=0, \qquad \varphi \in \mathcal H_{\rm ac}.
$$
Furthermore, the wave operators fulfill the important intertwining relations,
\beq\label{intertw}
f(H) \, P_{\rm ac }(H)= W_\pm(H,H_0) \,f(H_0)\, P_{\rm ac}(H_0) \, W_\pm(H,H_0)^\ast,
\ene
where $f$  is a Borel function. For these results see \cite{rs3}. The intertwining relations allow us to obtain important properties of $f(H) \, P_{\rm ac}(H)$ from those of
 $f(H_0) P_{\rm ac}(H_0).$ Let us explain. Assume that  the wave operators $W_\pm(H,H_0)$  are bounded in a Banach space, $Y,$  and the adjoints  $W_\pm(H,H_0)^\ast$  are bounded in a Banach space, $X.$  Then, if $f(H_0) P_{\rm ac}(H_0)$  is bounded from $X$ into $Y,$    it follows from \eqref{intertw}  that also $f(H) P_{\rm ac}(H)$ is bounded between the same spaces, and furthermore,
 \beq\label{equivalence}
   \|  f(H) P_{\rm ac}(H)  \|_{\mathcal B(X,Y)}\leq C \,  \| f(H_0) P_{\rm ac}(H_0) \|_{\mathcal B(X,Y)},
  \ene 
  for some constant $C$ and where $\mathcal B(X,Y)$ denotes the Banach space of bounded operators from $X$ into $Y.$ In the applications $H_0$ is often a constant coefficients operator and $f(H_0)$ is a Fourier multiplier. It is usually a simple matter to obtain important dispersive estimates, like the $L^p-L^{p'},$  estimates, $\frac{1}{p}+ \frac{1}{p'}=1, 1 \leq p \leq \infty,$  and the Strichartz estimates for the  free Schr\"odinger equation \eqref{fseq}
 with Hamiltonian $H_0,$ and then \eqref{equivalence}  gives us these estimates for the interacting Schr\"odinger equation \eqref{aschr} with Hamiltonian $H.$ These dispersive estimates play a crucial role in the study of initial value problems and in the scattering theory of nonlinear dispersive equations, like the nonlinear Schr\"odinger equation, and also in other problems, like the stability of soliton solutions. See \cite{Fanelli} and \cite{Schlag}.
 
 The wave operators are singular integral operators in the spectral representation of the unperturbed operator. On this point see \cite{fa64} and Section 1 in Chapter 4 of \cite{ya92}  where this question is discussed. As singular integral operators are bounded in $L^p$ spaces, the wave operators are bounded in $L^p$ spaces in the spectral representation of the unperturbed operator. I thank  D. R. Yafaev for calling this fact to my attention. Note  that in this paper we consider the related, but different, problem of the boundedness of the wave operators in $L^p$ spaces in the configuration representation.

The paper is organized as follows. In Section ~\ref{nota} we  introduce the notation that we use. In Section ~\ref{halfline} we state our results on the boundedness of the wave operators on the half line. In Section ~\ref{line} we state our results on the boundedness of the wave operators on the  line. In Section ~\ref{proofs} we mention  the results on the scattering theory of matrix Schr\"odinger equations that we need and we give the proofs of our theorems.

\section{Notation}\label{nota}\sss
We denote by $\mathbb{R}^{+}$ the positive  half line, $(0,\infty),$ and we designate by $\mathbb C$ the complex numbers. For a vector $Y \in \mathbb C^n$ we denote by $Y^T$ its transpose. By $\langle\cdot,\cdot\rangle$ we designate the scalar product in $\mathbb{C}^{n}.$  We introduce the following convenient notation. For any vector $  Y=(y_1,y_2,\dots, y_{2n})^T\in \mathbb C^{2n}$ we denote $Y_+:= (y_1,y_2,\dots, y_n)^T \in \mathbb C^n$ the vector with the first $n$ components of $Y,$ and $Y_-:=(y_{n+1}, y_{n+2}, \dots, y_{2n})^T\in \mathbb C^n$ the vector with the last $n$ components of $Y.$ Further, we use the notation $Y=(Y_+,Y_-)^T.$ 

We denote the entries of a $n\times m$ matrix $M$ by $ \{M\}_{i,j}, 1 \leq i \leq n, 1\leq j \leq m.$ By $0_{n}$ and $I_{n},$ $n=1,2,\dots,$ we designate the $n\times n$ zero and identity matrices, respectively. By $|M|$ we denote the norm of a matrix $M.$
We designate  by $L^{p}(U, \mathbb C^n)$, $1\leq  p \leq\infty,$ where
$U={\mathbb{R}}^{+}$ or $U={\mathbb{R}}$,
the Lebesgue spaces of $\mathbb{C}^{n}$ valued functions defined on $U.$ Let us denote by $C^\infty_0(U, \mathbb C^n)$  the space of all infinitely differentiable functions  defined on $U$ and that have compact support.
We designate  by  $ L^{p}(U,M_n)$,  $1\leq p \leq\infty,$ the  Lebesque space of $ n \times n$ matrix valued functions defined on $U.$ Further, we designate by $L^1_\gamma(U, M_n), \gamma > 0,$ the Lebesque space of $n \times n$ matrix valued functions defined on $U$ such that 
$$
\int_U\, (1+|x|)^\gamma\, |V(x)|\, dx < \infty.
$$
 For an integer $m \geq 1,$ $\mathbf H^{(m)}\left(  U, \mathbb C^n\right),$ where
$U={\mathbb{R}}^{+}$ or $U={\mathbb{R}}$,  is the
standard Sobolev space   of $\mathbb C^n$ valued functions (see \cite{AdamsF} for the definition and the properties of these spaces). By  $\mathbf H^{( m, 0)}\left( \mathbb R^+, \mathbb C^n\right), m \geq 1,$  we denote the  closure of $C_{0}^{\infty}\left( \mathbb R^+, \mathbb C^n\right)$ in the space $H^{m}\left( \mathbb R^+, \mathbb C^n\right)$. Note that the functions in  $\mathbf H^{\left( m, 0\right) }\left( \mathbb R^+,\mathbb C^n \right),$ as well as their derivatives of order up to $m-1,$ are zero at $x=0.$

 The Fourier transform, and the inverse Fourier transform are  designated by,
\[
\mathcal{F}f(k):=\ds \frac{1}{\sqrt{2 \pi}}\int_{\mathbb{R}}\,e^{-ikx}\,f(x)\,dx, \qquad 
\mathcal{F}^{-1}f(x):=\ds \frac{1}{\sqrt{2\pi}}\,\int_{\mathbb{R}}\,e^{ikx}\,f(k)\,dk.
\]
 
 For any set $ O \subset \mathbb R,$ we denote by $\chi_{ O}$ the characteristic function of $ O.$

For any operator $G$ in a Banach space $X$ we denote by
$D[G]$ the domain of $G.$ Further, for a densely defined operator $G$ in a Banach  space we denote by $G^\dagger$ its adjoint.  For any selfadjoint operator, $H,$ in a Hilbert space  and  for any Borel set $O$  we designate by  $E(O; H) $  the
spectral projector of $H$ for $O$.  
 
%
We designate by $\mathcal E_{\text{\rm even}}$ the extension operator from  $L^p(\mathbb R^+, \mathbb C^n), 1 \leq p \leq \infty,$ to even functions in $L^p(\mathbb R, \mathbb C^n)$  as follows
$$
(\mathcal E_{\text{\rm even}} Y)(x):=
\begin{cases}
Y(x), \qquad  x >0,\\
Y(-x), \qquad x \leq 0.
\end{cases}
$$
Clearly, $\mathcal E_{\text{\rm  even}}$ is bounded from $L^p(\mathbb R^+, \mathbb C^n) $ into  $L^p(\mathbb R, \mathbb C^n), 1 \leq p \leq \infty.$ 

Moreover, we denote by $\mathcal E_{\text{\rm odd}}$ the extension operator from  $L^p(\mathbb R^+, \mathbb C^n), 1 \leq p \leq\infty,$ to odd functions in $L^p(\mathbb R, \mathbb C^n)$  in the following way
$$
(\mathcal E_{\text{\rm odd}} Y)(x):=
\begin{cases}
Y(x), \qquad  x >0,\\
-Y(-x), \qquad x \leq 0.
\end{cases}
$$
We have that $\mathcal E_{\text{\rm  odd}}$ is bounded from $L^p(\mathbb R^+, \mathbb C^n),$ into   $L^p(\mathbb R, \mathbb C^n), 1 \leq p \leq \infty.$ 

We denote by $\mathcal R$ the restriction operator from $ L^p(\mathbb R, \mathbb C^n)$ into $L^p(\mathbb R^+, \mathbb C^n), 1 \leq p \leq \infty,$ given by,
$$
(\mathcal R Y)(x):= Y(x), \qquad x >0.
$$
We have that $\mathcal R$ is bounded from $L^p(\mathbb R, \mathbb C^n),$ into $L^p( \mathbb R^+, \mathbb C^n), 1 \leq p \leq  \infty.$

   For any integrable  $n \times n$ matrix valued function, $G(x), x \in \mathbb R,$  we denote by $Q(G)$ the operator of convolution by $ G( x),$
$$
\left(Q(G)Y\right)(x):= \int_{\mathbb R}\, G(x-y)\, Y(y)\, dy =  \int_{\mathbb R}\, G( y)\, Y(x-y)\, dy.
$$
Since $G$ is integrable, the operator  $Q(G)$ is bounded in $L^p(\mathbb R, \mathbb C^n), 1 \leq p \leq \infty.$ For any $n\times n$ matrix valued measurable function $K(x,y)$ defined for $x,y \in \mathbb R^+,$
 we denote by $\mathbf K(K)$ the operator,
$$
{{\mathbf K}(K) Y}(x):= \int_{\mathbb R^+}\, K(x,y)\, Y(y)\, dy.
$$
The operator ${\mathbf K}(K)$ is bounded in $L^p(\mathbb R^+, \mathbb C^n), 1 \leq p \leq \infty,$ provided that the following two conditions are satisfied,
\beq\label{w.13.c}
\sup_{x \in \mathbb R^+}\, \int_{ \mathbb R^+}\, |K(x,y)|\, dy < \infty,\qquad \sup_{y \in \mathbb R^+}\, \int_{\mathbb R^+}\, |K(x,y)|\, dx <\infty.
\ene

The Hilbert transform, $\mathcal H,$ is defined as follows,
$$
(\mathcal H Y)(x):= \frac{1}{\pi}\, \text{\rm PV}\, \int_{\mathbb R}\, \frac{Y(y)}{x-y}\, dy,
$$
where $\text{\rm PV}$ means the principal value of the integral. As is well  known \cite{sado,stein}, the Hilbert transform is a bounded operator in $L^p(\mathbb R, \mathbb C^n), 1 < p <\infty.$

\section{The wave operators on the half line}\label{halfline}\sss
To define the wave operators we take as unperturbed Hamiltonian $ H_0$ the selfadjoint realization in $L^2(\mathbb R^+, \mathbb C^n)$ of the formal differential  operator $\ds-\frac{d^2}{dx^2}$  with  the Neumann boundary condition, $\ds \frac{d}{dx}Y(0)=0,$ see Section ~\ref{proofs} below and  Sections 3.3 and 3.5 of Chapter three   of  \cite{WederBook}.  This choice is motivated by the theory of quantum graphs  \cite{Kostrykin1,Kostrykin2}.
Note that the spectrum of $H_0$  is absolutely continuous and  that it coincides with $[0, \infty).$    
The perturbed  Hamiltonian, $H,$
is the selfadjoint  realization  in $L^{2}\left(  \mathbb{R}^{+},\mathbb C^n\right)  $  of the formal differential operator $\ds- \frac{d^2}{dx^2}+V(x)$ with the boundary condition
\beq\label{MSA}
-B^{\dagger}Y\left(0\right)  +A^{\dagger} \frac{d}{dx}Y(0)=0,
\ene
 where the constant matrices $A,B$ satisfy \eqref{wcon1} and \eqref{wcon2}, and the potential $V$ fulfills  \eqref{PotentialHermitian} and \eqref{PotentialL1}. For the definition of $H$ see Section ~\ref{proofs} below  and  Sections 3.3 and 3.5 of Chapter three of \cite{WederBook}.

The wave operators, $W_\pm(H,  H_0 ),$ are defined as follows,
\beq\label{w.2}
W_\pm(H, H_0 ):= \text{\rm s-} \lim_{t \rightarrow \pm \infty} e^{i t H}\, e^{-it H_0},
\ene
since $P_{\rm ac}(H_0)=I.$ 
It is proven  in Section 4.4 of Chapter four of \cite{WederBook}  that the  wave operators $ W_\pm(H, H_0 )$ exist, and  are complete.

Our result in the case of $L^p(\mathbb R^+, \mathbb C^n), 1 < p < \infty,$ is the following theorem.

\begin{theorem}\label{bounded}
Suppose that $V$ fulfills \eqref{PotentialHermitian} and \eqref{PotentialL11} and that the constant  matrices $A,B$ satisfy \eqref{wcon1}, and \eqref{wcon2}.  Then, for all $Y \in L^2(\mathbb R^+, \mathbb C^n)$ we have,
\beq\label{w.14}
W_\pm(H,H_0)Y = \sum_{j=1}^3\, W_\pm^{(j)}\,Y,
\ene
where,
\beq\label{w.15}
W_\pm^{(1)}Y:=(I+ {\mathbf K}(K))\mathcal R \left( \ds \frac{\pm i}{2}\, \mathcal H \mathcal E_{\text{\rm even}} Y+ \ds\frac{1}{2}\,\mathcal E_{\text{\rm even}}Y\right),
\ene
\beq\label{w.16}
W_\pm^{(2)}Y:= (I+ {\mathbf K}(K)) \mathcal R \left(\ds \frac{\mp i}{2}\, (\mathcal H  S_\infty \mathcal E _{\text{\rm even}} Y)+ \ds\frac{1}{2}\,S_\infty\mathcal  E_{\text{\rm even}} Y\right),
\ene
\beq\label{w.17}
W_\pm^{(3)}Y:=(I+ {\mathbf K}(K)) \mathcal R \left(\ds \frac{\mp i}{2}\, (\mathcal H Q (F_s)  \mathcal E_{\text{\rm even}} Y)+\ds \frac{1}{2} (Q(F_s) \mathcal E_{\text{\rm even}} Y)\right).
\ene
Furthermore, the wave operators $W_\pm(H,H_0)$ restricted to  $L^2(\mathbb R^+, \mathbb C^n)\cap L^p(\mathbb R^+, \mathbb C^n), 1 <  p < \infty,$ extend uniquely to bounded  operators in $L^p(\mathbb R^+, \mathbb C^n), 1 < p < \infty$ and equations \eqref{w.14}-\eqref{w.17} hold for all $ Y \in L^p(\mathbb R^+, \mathbb C^n), 1 < p < \infty.$ Moreover, the adjoints of the wave operators $W_\pm(H,H_0)^\dagger,$ restricted to  $L^2(\mathbb R^+, \mathbb C^n)\cap L^p(\mathbb R^+, \mathbb C^n), 1 <  p < \infty,$ extend uniquely to bounded  operators on $L^p(\mathbb R^+, \mathbb C^n), 1 < p < \infty.$ The $n\times n$ matrix valued function $K(x,y), x, y \in \mathbb R^+$ is defined in  \eqref{marchenko}.  Moreover, the quantity $S_\infty$ is defined in  \eqref{sinft}, and the $n \times n$ matrix valued function $F_s(x), x \in \mathbb R$ is defined in \eqref{fs}. 
\end{theorem}
Our result in the case of $L^1(\mathbb R^+, \mathbb C^n)$ and in $L^\infty(\mathbb R^+, \mathbb C^n)$ is stated in the next  theorem. We first prepare a convenient notation, where the scattering matrix $S(k)$ is defined in \eqref{Scatteringmatrix}.

\beq\label{new.0000}
P_+(x):=  \frac{1}{\sqrt{2\pi}}\,\left(\mathcal F\,  \chi_{\mathbf R^+}(k)\, (S(-k)-S_\infty)\right)(x), P_-(x):=  \frac{1}{\sqrt{2\pi}}\,\left(\mathcal F^{-1}\,  \chi_{\mathbf R^+}(k)\, (S(k)-S_\infty)\right)(x).
\ene

\begin{theorem} \label{theo.new.1} 
  Suppose that $V$ fulfills \eqref{PotentialHermitian}, that $V \in L^1_\gamma(\mathbb R^+,M_n), \gamma > \frac{5}{2},$ that the constant  matrices $A,B$ satisfy \eqref{wcon1}, and \eqref{wcon2}, and that $S(0)= S_\infty=I_n$.  Then, for all $Y \in L^2(\mathbb R^+, \mathbb C^n)$ we have,
\beq\label{new.000.aa}
W_\pm(H,H_0)Y = Y+\mathbf K(K) Y+\mathcal R Q(P_\pm)\mathcal E_{\rm even}Y+ \mathbf K(K)\mathcal R\, Q(P_\pm)\mathcal E_{\rm even}Y.
\ene
Furthermore, the wave operators $W_\pm(H,H_0)$ and $W_\pm(H,H_0)^\dagger$ restricted to $L^2(\mathbb R^+)\cap L^1(\mathbb R^+)$, respectively to $L^2(\mathbb R^+)\cap L^\infty(\mathbb R^+),$ extend to bounded operators on $L^1(\mathbb R^+)$ and to bounded operators on $L^\infty(\mathbb R).$
 The $n\times n$ matrix valued function $K(x,y), x, y \in \mathbb R^+$ is defined in  \eqref{marchenko}.  Moreover, the scattering matrix, $S(k), k \in \mathbb R,$ is defined in \eqref{Scatteringmatrix}, the quantity $S_\infty$ is defined in  \eqref{sinft}, and the $n \times n$ matrix valued functions $P_\pm(x)$ are defined in \eqref{new.0000}.
\end{theorem}

In Remark~\ref{rem.new2} we prove by means of a counter example that the condition $S(0)= S_\infty=I_n$ is necessary for the boundedness of the wave operators on $L^1(\mathbb R^+)$ and on $L^\infty(\mathbb R).$

 \begin{remark} \label{rem.new3}{\rm
 It follows from equation (3.10.37) in page 197 of \cite{WederBook} that $S_\infty=I_n$ if and only if there are no Dirichlet boundary conditions in the diagonal representation of the boundary matrices given in \eqref{A,Btilde}, \eqref{PSM13}, and \eqref{trans}. Further, by Theorem 3.8.13 in page 137 and Theorem 3.8.14 in pages 138-139 of \cite{WederBook}, $S(0)=I_n$ if and only if the geometric multiplicity, $\mu,$ of the eigenvalue zero of the zero energy Jost matrix, $J(0),$  (see \eqref{Jostmatrix}) is equal to $n.$ Moreover, by Remark 3.8.10 in pages 129-130 of \cite{WederBook} the geometric multiplicity of the eigenvalue zero of $J(0)$ is equal to $n$ if and only if there are $n$ linearly independent bounded solutions to the Schr\"odinger equation \eqref{MSStationary}with zero energy, $k^2=0,$ that satisfy the boundary condition \eqref{MSA}. This corresponds to the { \it purely exceptional case} where there are $n$ linearly independent half-bound states or zero-energy resonances.  We provide below a simple example of this situation, with a non-trivial potential. Consider the scalar case, $n=1,$ with the potential,
 $$
 V(x)= \left\{\begin{array}{l} 0, \,\,x  > 1, \\
 1, \,\,0 < x <1.
 \end{array}
 \right.
 $$
 The Jost solution (see \eqref{Jostsolution}) is computed in  Example 6.4.1 in pages 536-538 of \cite{WederBook}. It is given by,
 $$
 f(k,x)= \left\{ \begin{array}{l} e^{ikx}, \qquad x \geq 1, \\
 \frac{1}{2}\, \left( 1+ \frac{k}{\gamma}  \right)\, e^{ik}\,e^{i\gamma (x-1)}+  \frac{1}{2}\, \left( 1-\frac{k}{\gamma}  \right)\, e^{ik}\,e^{-i\gamma (x-1)},\qquad 0 \leq x \leq 1. 
 \end{array}
 \right.
 $$
 We take the boundary matrices, $ A=- \sin\theta, B=\cos\theta,$ with $\theta= \arctan \coth 1.$ The boundary condition is
$\cos\theta Y(0)+\sin\theta Y'(0)=0.$ The Jost function is given by, $ J(k)= f(k,0)\cos\theta+ f'(k,0)\sin\theta.$ We have, $J(0)=0. $
Then $S(0)=1,$ and as we have the Robin boundary condition, $S_\infty=1.$ Of course, these results can be obtained by explicit computation.}
  \end{remark}

\section{The wave operators on  the line}\label{line}\sss
We obtain our results on the line proving  that a $2n\times2n$ matrix Schr\"{o}dinger
equation on the half line is unitarily equivalent to a $n\times n$ matrix
Schr\"{o}dinger equation on the line with a point interaction at $x=0.$ For this purpose we follow  Section 2.4 in Chapter 2 of \cite{WederBook}.
Let us denote by  $\mathbf{U}$ the unitary operator from $L^{2}\left(  \mathbb{R}
^{+},\mathbb{C}^{2n}\right) $ onto $L^{2}\left(  \mathbb{R}, \mathbb{C}^n\right),$ defined as follows,
\begin{equation}\label{unitary}
Y\left(  x\right)  =\mathbf{U }Z\left(  x\right)  :=\left\{
\begin{array}
[c]{c}%
Z_{+}\left(  x\right)  ,\text{ \ }x\geq0,\\
Z_{-}\left(  -x\right)  ,\text{ \ }x<0,
\end{array}
\right.  %
\end{equation}
where $Z=\left(  Z_{+},Z_{-}\right)  ^{T},$
 with $Z_{+}, Z_-\in L^{2}\left(
\mathbb{R}^{+},\mathbb{C}^{n}\right).$ Let us take as potential   the diagonal matrix
$$
V(x):=\left\{
\begin{array}
[c]{lc}%
V_+(x) &  0_n\\
0_n& V_-(x)%
\end{array}
\right\}, 
$$

where $V_{+}, V_-$ are selfadjoint $n \times n$ matrix-valued functions that belong to $L^{1}_{1}(\mathbb R^+, M_n).$
 Under the action of the unitary transformation  $\mathbf{U}$ the
Hamiltonian in the half line, $H,$ is  unitarily transformed into the Hamiltonian on the line, $H_\mathbb R,$ as follows,
\begin{equation}\label{hwhole}
H_{\mathbb{R}}:=\mathbf{U}\,H\mathbf{U}^{\dagger}, \qquad
 D[H_{\mathbb{R}%
}]:=\{Y\in L^{2}\left(  \mathbb{R},\mathbb{C}^{n}\right)  :\mathbf{U}%
^{\dagger}Y\in D[H]\}.
\ene
The operator $H_{\mathbb{R}}$ is a selfadjoint realization in $L^{2}\left(
\mathbb{R},\mathbb{C}^{n}\right)  $ of the formal differential operator
$\ds-\frac{d^{2}}{d x^2}+\mathcal V(x)$ where the  selfadjoint $n\times n$ matrix valued potential, $\mathcal V,$ is given by,
\[
\mathcal V\left(  x\right)  =\left\{
\begin{array}
[c]{c}%
V_{+}\left(  x\right)  ,\text{ \ }x\geq0,\\
V_{-}\left(  -x\right)  ,\text{ \ }x<0.
\end{array}
\right.
\]
Note that $ \mathcal  V \in L^1_1(\mathbb R, M_n).$ The boundary condition \eqref{MSA} satisfied by the functions in the domain of $H$ implies that the functions in the domain of $H_\mathbb R$ fulfill a transmission condition at $x=0.$ To compute this transmission condition it is convenient to write the matrices $A,B$ in \eqref{MSA} in the following way,
\begin{equation}
A=\left\{
\begin{array}
[c]{l}%
A_{1}\\
A_{2}%
\end{array}
\right\}  ,\quad  B= \left\{
\begin{array}
[c]{l}%
B_{1}\\
B_{2}%
\end{array}
\right\}  , \label{matrices}%
\end{equation}
where $A_{j},B_{j},j=1,2,$  are $n\times2n$ matrices. Hence, \eqref{MSA} implies  that the
functions in the domain of $H_{\mathbb{R}}$ satisfy the following transmission
condition at $x=0,$%

\begin{equation}
-B_{1}^{\dagger}Y(0^+)-B_{2}^{\dagger}Y(0^-)+A_{1}^{\dagger}\frac{d}{dx}Y(0^+)-A_{2}^{\dagger}\frac{d}{dx}Y(0^-)=0. \label{bdcond}%
\end{equation}
Remark that  $u(t,x)$ is a solution of the problem \eqref{0.0},  (\ref{MS}) if and only if
$v(t,x):=\mathbf{U}u(t,x)$ is a solution of the following $n\times n$ matrix equation
on the line,
\begin{equation}
\left\{
\begin{array}
[c]{c}%
\ds i\frac{\partial}{\partial t} v\left(  t,x\right)  =\left(  -\frac{\partial^2}{\partial x^2}+\mathcal V\left(
x\right)  \right)  v\left(  t,x\right)  ,\ t\in\mathbb{R},\,x\in\mathbb{R},\\
v\left(  0,x\right)  =v_{0}\left(  x\right)  :=\mathbf{U}u_{0}\left(
x\right)  ,x\in\mathbb{R},\\
-B_{1}^{\dagger}v(t,0^+)-B_{2}^{\dagger}v(t,0^-)+A_{1}^{\dagger}\ds \frac{\partial}{\partial x}v(t,0^+)-A_{2}^{\dagger}\ds\frac{\partial}{\partial x}v(t,0^-)=0.
\end{array}
\right.  \label{MSR}%
\end{equation}
Below we give  an example. Let $A,B$ be the following matrices, 
\beq\label{matrices2}
A=\left\{
\begin{array}
[c]{lc}%
0_{n} & I_{n}\\
0_{n} & I_{n}%
\end{array}
\right\}  ,\quad B=\left\{
\begin{array}
[c]{lc}%
-I_{n} & \Lambda\\
\ I_{n} & 0_{n}%
\end{array}
\right\}  ,
\ene
where $\Lambda$ is a selfadjoint $n\times n$ matrix. It is  easy to check that  these  matrices satisfy 
(\ref{wcon1}, \ref{wcon2}). The transmission condition in
(\ref{MSR}) is given by,
\begin{equation}
v(t,0^+)=v(t,0^-)=v(t,0),\quad\ds \frac{\partial}{\partial x}v(t,0^+)-\frac{\partial}{\partial x}%
v(t,0^-)=\Lambda v(t,0). \label{bcond2}%
\end{equation}
This transmission condition is  a Dirac-delta point interaction at
$x=0$ with coupling matrix $\Lambda$.  In the particular case $\Lambda=0,$  the functions   $v(t,x)$ and
$ \frac{\partial}{\partial x}v(t,x)$ are continuous at $x=0$ and we have
 the matrix Schr\"{o}dinger equation on the line without a
point interaction at $x=0.$ 

Let us denote by $H_{0,\mathbb R}$ the Hamiltonian \eqref{hwhole} with the potential $\mathcal V$ identically zero and with the boundary condition given by the matrices \eqref{matrices2} with $\Lambda=0.$ Note that $H_{0,\mathbb R}$ is the standard selfadjoint realization of the formal  differential operator $ -\frac{d^2}{dx^2}$ with domain,
$
D[H_{0,\mathbb R}]:= \mathbf H^{(2)}(\mathbb R, \mathbb C^n).
$
In particular, $H_{0,\mathbb R}$ is absolutely continuous and its spectrum consists of  $[0,\infty).$ We define the wave operators on the line as follows,
\beq\label{zzz.1}
W_\pm(H_\mathbb R, H_{0,\mathbb R}):= \text{\rm s-} \lim_{t \to \pm \infty}\, e^{it H_\mathbb R}\, e^{-it H_{0,\mathbb R}}. 
\ene
Using Theorem ~\ref{bounded} and the unitary transformation \eqref{hwhole} we prove the  following theorem, on the boundedness of the wave operators on $L^p(\mathbb R, \mathbb C^n), 1 < p < \infty.$
\begin{theorem}
\label{wholeline}  Let $H_\mathbb R$ be the Hamiltonian \eqref{hwhole}, with the transmission condition \eqref{bdcond}, and where $\mathcal V\left(  x\right),$ $x\in\mathbb{R}$, is a  $n\times n$ selfadjoint matrix-valued function, i.e., $\mathcal V(x)=\mathcal V^\dagger(x)$  and, moreover, $\mathcal V$ satisfies \eqref{ddddd}. 
 Then, the wave operators $W_\pm(H_\mathbb R, H_{0,\mathbb R})$ exist and are complete. Moreover, the $W_\pm(H_\mathbb R, H_{0,\mathbb R}),$ and the adjoint wave operators  $W(H_\mathbb R,H_{0,\mathbb R})^\dagger,$  restricted to  $L^2(\mathbb R, \mathbb C^n)\cap L^p(\mathbb R, \mathbb C^n), 1 <  p < \infty,$ extend uniquely to bounded  operators in $L^p(\mathbb R, \mathbb C^n), 1 < p < \infty.$
\end{theorem}
Theorem~ \ref{wholeline} generalizes the results obtained in  \cite{we-99,gy,daf,dmw} to  the case of general point interactions at $x=0,$ and to potentials that satisfy \eqref{pscalar} with $ \gamma=1.$

Below we state our theorem on the boundedness  of the wave operators on the line in $L^1(\mathbb R, \mathbb C^n),$ and in $L^\infty(\mathbb R, \mathbb C^n).$

 \begin{theorem} \label{theo.new.6} 
 Let $H_\mathbb R$ be the Hamiltonian \eqref{hwhole}, with the transmission condition \eqref{bdcond}, and where $\mathcal V\left(  x\right),$ $x\in\mathbb{R}$, is a  $n\times n$ selfadjoint matrix-valued function, i.e., $\mathcal V(x)=\mathcal V^\dagger(x),$ and moreover,  $\mathcal V \in L^1_\gamma(\mathbb R, M_n), \gamma > \frac{5}{2}.$ Assume that
  $S_{ \mathbb R}(0)= S_{\mathbb R, \infty}= I_{2n}.$
  Then, the wave operators $W_\pm(H_{\mathbb R},H_{0,\mathbb R})$ and $W_\pm(H_{\mathbb R},H_{0,\mathbb R})^\dagger$ restricted to $L^2(\mathbb R,\mathbb C^n)\cap L^1(\mathbb R,\mathbb C^n)$, respectively to $L^2(\mathbb R,\mathbb C^n)\cap L^\infty(\mathbb R,\mathbb C^n),$ extend to bounded operators on $L^1(\mathbb R,\mathbb C^n)$ and to bounded operators on $L^\infty(\mathbb R,\mathbb C^n).$ The scattering matrix on the line, $S_{\mathbb R}(k), k \in \mathbb R,$ is defined in \eqref{new.55.aa} and the quantity $S_{\mathbb R, \infty}$ in \eqref{new.57}.
\end{theorem}

\begin{remark} \label{re.new7}{\rm
In the case where the matrices $A,B$ are equal to the matrices in  \eqref{matrices2} and $\Lambda=0,$  there is  no point interaction at $x=0.$ The scattering theory in this situation has been studied in \cite{akv}, and the references quoted there. 
In this case $S_{ \mathbb R}(0)= S_{\mathbb R, \infty}= I_{2n}$ in the {\it purely exceptional case} where there are $n$ linearly independent bounded solutions to the Schr\"odinger equation \eqref{seline} with zero energy, $k^2=0$ (in this case we say that there are $n$ linearly independent zero-energy resonances, or half-bound states), and if moreover,  the zero-energy Jost solution from the left, $f_l(0,x),$ satisfies,
$
\lim_{x \to -\infty} f_l(0,x)=I_n
$
For the definition of $f_l(k,x), k \in \mathbb R,$ see \eqref{new.41}. 

As mentioned above, in Theorem~1.1 of  \cite{we-99} we proved that in the scalar case, $n=1,$ and without  point interactions,  the wave operators $W_\pm(H_{\mathbb R},H_{0,\mathbb R})$ and $W_\pm(H_{\mathbb R},H_{0,\mathbb R})^\dagger$  extend to bounded operators on $L^1(\mathbb R,\mathbb C)$ and to bounded operators on $L^\infty(\mathbb R,\mathbb C)$ in the { exceptional case}, and  assuming that $ \lim_{x \to -\infty} f_l(0,x)=1$, for potentials that satisfy \eqref{pscalar} with  $
 \gamma > \frac{5}{2}.$ Theorem~\ref{theo.new.6} generalizes this result of \cite{we-99} to the case where there is a general point interaction.}
\end{remark}

\section{ Scattering theory and the $L^p$- boundedness of the wave operators\label{proofs}}
\sss
\subsection{Scattering theory for the  matrix Schr\"{o}dinger equation  on the half line.}

 We study the following  the stationary matrix Schr\"{o}dinger equation
on the half line%
\begin{equation}
-\frac{d^2}{d x^2}Y(x)+V( x) Y(x)=k^{2}Y(x),\text{ }x\in
\mathbb{R}^{+}. \label{MSStationary}%
\end{equation}
In this equation  $k^{2}$ is the complex-valued spectral parameter,  the $n \times n$ matrix  valued potential $V(x)$ satisfies
(\ref{PotentialHermitian}) and moreover,
\begin{equation}
V\in L^{1}\left(  \mathbb{R}^{+}, M_n\right). \label{PotentialL1}%
\end{equation}
The solution  $Y$ that  appears in \eqref{MSStationary} is  either  a column vector with $n$ components, or a $n\times n$
matrix-valued function.
As we already 
mentioned, the  general selfadjoint
boundary condition at $x=0$ can be expressed in terms of two constant $n\times
n$ matrices $A$ and $B$ as in \eqref{MSA},
where the matrices $A$ and $B$ fulfill \eqref{wcon1}, \eqref{wcon2}.

Actually, there is a simpler equivalent  form  of the boundary condition \eqref{MSA}. In fact, in   \cite{[9]}, and in Section 3.4 of Chapter three of \cite{WederBook}, it is  given the explicit steps to go from any pair of
matrices $A$ and $B$ appearing in the selfadjoint boundary condition \eqref{MSA}, and that satisfy 
\eqref{wcon1},and  (\ref{wcon2}) to a pair $\tilde{A}$ and $\tilde{B},$ given by%
\begin{equation}
\tilde{A}=-\operatorname*{diag}\left\{\sin\theta_{1},...,\sin\theta_{n}\right\},\text{
}\tilde{B}=\operatorname*{diag}\left\{\cos\theta_{1},...,\cos\theta_{n}\right\},
\label{A,Btilde}%
\end{equation}
with appropriate real parameters $\theta_{j}\in(0,\pi].$ The matrices  $\tilde{A}, \tilde{B}$   satisfy
(\ref{wcon1}), (\ref{wcon2}). In the case of  the matrices $\tilde{A}$, $\tilde{B},$ the
boundary condition (\ref{MSA}) is given by%
\begin{equation}
\cos\theta_{j}Y_{j}\left(  0\right)  +\sin\theta_{j}\frac{d}{d x}Y_{j}
\left(  0\right)  =0,\text{ \ \ }j=1,2,...,n. \label{PSM13}%
\end{equation}
The case $\theta_{j}=\pi$
corresponds to the Dirichlet boundary condition and the case $\theta_{j}%
=\pi/2$ corresponds to the Neumann boundary condition. In the general case, there are
$n_{\operatorname*{N}}\leq n$ values with $\theta_{j}=\pi/2$ and
$n_{\operatorname*{D}}\leq n$ values with $\theta_{j}=\pi$. Further, there
are $n_{\operatorname*{M}}$ remaining values, where $n_{\operatorname*{M}%
}=n-n_{\operatorname*{N}}-n_{\operatorname*{D}}$ such that those $\theta_{j}%
$-values lie in the interval $(0,\pi/2)$ or $(\pi/2,\pi).$
  It is proven  in \cite{[9]}, and in Section 3.4 of Chapter three of \cite{WederBook},   that for any
pair of matrices $(A,B)$ that satisfy (\ref{wcon1}, \ref{wcon2}) there is a pair
of matrices $(\tilde{A},\tilde{B})$ as in (\ref{A,Btilde}), a unitary matrix
$M$ and two invertible matrices $T_{1},T_{2}$ such%

\begin{equation}
A=M\,\tilde{A}T_{1}M^{\dagger}T_{2},\quad B=M\,\tilde{B}T_{1}M^{\dagger}T_{2}.
\label{trans}%
\end{equation}
 As we will see, the Hamiltonians with the boundary condition given by matrices $A,B$ and with the matices $\tilde{A}, \tilde{B},$ are unitarily equivalent.

We construct a selfadjoint realization of the matrix Schr\"{o}dinger operator
$-\frac{d^2}{dx^2}+V(x)$ by quadratic forms methods. For the following
discussion see Sections 3.3 and 3.5 of Chapter three in \cite{WederBook}. Let $\theta_{j}$ be as in  equations
(\ref{A,Btilde}). We denote
\begin{equation}
\widehat{\mathbf H}_{j}^{(1)}(\mathbb R^+, \mathbb C):=\mathbf H^{\left(1, 0\right)  }(\mathbb R^+, \mathbb C),\text{ if }\theta_{j}%
=\pi,\text{ \rm and }\widehat{\mathbf H}_{j}^{(1)}(\mathbb R^+,\mathbb C):=\mathbf H^{(1)}(\mathbb R^+, \mathbb C),\text{ if }\theta_{j}\neq\pi.
\label{spaceW1j}%
\end{equation}
We put%
\[
\widetilde{\mathbf H}^{(1)}(\mathbb R^+, \mathbb C^n):=\oplus_{j=1}^{n}\widehat{\mathbf H}_{j}^{(1)}(\mathbb R^+,\mathbb C).
\]
We define%
\[
\Theta:=\operatorname*{diag}[\widehat{\cot}\theta_{1},...,\widehat{\cot}%
\theta_{n}],
\]
where $\widehat{\cot}\theta_{j}=0,$ if $\theta_{j}=\pi/2,$ or $\theta_{j}%
=\pi,$ and $\widehat{\cot}\theta_{j}=\cot\theta_{j},$ if $\theta_{j}\neq
\pi/2,\pi.$ Suppose that the potential $V$ satisfies (\ref{PotentialHermitian}%
) and (\ref{PotentialL1}).
The following quadratic form is closed, symmetric and bounded below,
\begin{equation}\begin{array}{l}
h\left(  Y,Z\right)  :=\left(  \frac{d}{dx}Y,\frac{d}{dx}Z\right)_{L^2(\mathbb R^+, \mathbb C^n)}
-\left\langle M\Theta M^{\dag}Y\left(  0\right)  ,Z\left(
0\right)  \right\rangle +\left(  VY,Z\right)_{L^2(\mathbb R^+, \mathbb C^n)},\\Q\left(
h\right)  :=\mathbf H^{(A,B)}(\mathbb R^+,\mathbb C^n), \end{array}\label{quadratic}%
\end{equation}
where by $Q(h)$ we denote the domain of $h$ and,
\begin{equation}
\mathbf H^{(A,B)}(\mathbb R^+,\mathbb C^n):=M\widetilde{\mathbf H}^{(1)}(\mathbb R^+, \mathbb C^n)\subset \mathbf H^{(1)}(\mathbb R^+, \mathbb C^n). \label{W1AB}%
\end{equation}
 We denote by $H_{A,B,V}$ the selfadjoint bounded below
operator associated to $h$ \cite{kato}. The operator $H_{A,B,V}$ is the
selfadjoint realization of $\ds-\frac{d^2}{dx^2}+V\left(  x\right)  $ with the
selfadjoint boundary condition (\ref{MSA}). When there is no possibility of
misunderstanding we will use the notation $H,$ i.e., $H\equiv H_{A,B,V}.$ It
is proven in Section 3.6 of Chapter three of \cite{WederBook} that,
\begin{equation}
H_{A,B,V}=MH_{\tilde{A},\tilde{B},M^{\dagger}VM}M^{\dagger}.
\label{diagonalization}%
\end{equation}

In the next proposition we introduce the Jost solution given in   \cite{AgrMarch}. See also  Sections 3.1 and 3.2 of \cite{WederBook}.
\begin{prop}
\label{PJostsolution}Suppose that the potential $V$ satisfies
 (\ref{PotentialL1}). For each fixed
$k\in\overline{\mathbb{C}^{+}}\backslash\{0\}$ there exists a unique $n\times
n$ matrix-valued Jost solution $f\left(  k,x\right)  $ to equation
(\ref{MSStationary}) satisfying the asymptotic condition%
\begin{equation}
f\left(  k,x\right)  =e^{ikx}\left(  I+o\left(1\right)  \right)
,\qquad x\rightarrow+\infty. \label{Jostsolution}%
\end{equation}
Moreover, for any fixed $x\in\lbrack0,\infty),$ $f\left(  k,x\right)  $ is
analytic in $k\in\mathbb{C}^{+}$ and continuous in $k\in\overline
{\mathbb{C}^{+}}\setminus\{0\}$. If, moreover, $V$ satisfies \eqref{PotentialL11} the Jost solution also exists at $k=0,$ and for each  fixed $x\in\lbrack0,\infty),$ $f\left(  k,x\right)  $ is continuous in $k\in\overline
{\mathbb{C}^{+}}.$ Furthermore if \eqref{PotentialL11} holds, for $k\in\overline{\mathbb{C}^{+}}\backslash\{0\},$ the $o(1)$ in \eqref{Jostsolution} can be replaced by $o\left(\frac{1}{x}\right).$
\end{prop}
Using the Jost solution  and  the boundary matrices $A$ and $B$ satisfying (\ref{wcon1})-(\ref{wcon2}),
  we construct the Jost matrix $J\left(  k\right),$
\begin{equation}
J\left(  k\right)  =f\left(  -k^*,0\right)  ^{\dagger}B-f^{\prime}\left(-k^*
,0\right)  ^{\dagger}A,\text{ \ }k\in\overline{\mathbb{C}^{+}},
\label{Jostmatrix}%
\end{equation}
where the asterisk denotes complex conjugation. For the following result see  \cite{AgrMarch}, and  also Theorem 3.8.1 in page 114 of \cite{WederBook}.

\begin{prop}
\label{Jostnozero}Suppose that the potential $V$ satisfies
(\ref{PotentialHermitian}) and (\ref{PotentialL11}). Then, the Jost matrix
$J\left(  k\right)  $ is analytic for $k\in\mathbb{C}^{+}$, continuous for
$k\in\overline{\mathbb{C}^{+}}$ and invertible for $k\in\mathbb{R}%
\diagdown\{0\}.$
\end{prop}

Let  $K\left(  x,y\right)  $ be defined as follows
\beq\label{marchenko}
K\left(  x,y\right)  =\left(  2\pi\right)  ^{-1}\int_{-\infty}^{\infty
}[f\left(  k,x\right)  -e^{ikx}I]e^{-iky}dk, \qquad  x, y \geq0.
\ene
We introduce the following quantities,
\[
\sigma\left(  x\right)  =\int_{x}^{\infty}\left\vert V\left(  y\right)
\right\vert dy,\text{ \ }\sigma_{1}\left(  x\right)  =\int_{x}^{\infty
}y\left\vert V\left(  y\right)  \right\vert dy,\text{ }x\geq0.
\]
Remark  that for potentials satisfying (\ref{PotentialL11}), both
$\sigma\left(  0\right)  $ and $\sigma_{1}\left(  0\right)  $ are finite, and
furthermore, $ \linebreak 
\int_{0}^{\infty}\,\sigma(x)\,dx=\sigma_{1}(0)<\infty.$

The following  proposition is given in  \cite{AgrMarch}. See also Proposition 3.28 in pages 77-78 of \cite{WederBook}.

\begin{prop}
\label{PK}Suppose that the potential $V$ satisfies (\ref{PotentialHermitian})
and (\ref{PotentialL11}). Then, we have.

\begin{enumerate}
\item
The matrix $K\left(  x,y\right)  $ is
continuous in $(x,y)$ in the region $0\leq x\leq y,$ and is
related to the potential via
\[
K\left(  x,x^+\right)  =\frac{1}{2}\int_{x}^{\infty}V\left(  z\right)  dz,\text{
\ }x\in\lbrack0,+\infty).
\]
\item
The matrix $K\left(  x,y\right)  $ satisfies,
\begin{align}
K\left(  x,y\right)   &  =0,\text{ }y<x, x,y \in [0,\infty),  \nonumber\\
\left\vert K\left(  x,y\right)  \right\vert  &  \leq\frac{1}{2}e^{\sigma
_{1}\left(  x\right)  }\sigma\left(  \frac{x+y}{2}\right)  ,\text{\ }
x,y,\in \mathbb R^+. \label{ESTK}%
\end{align}%
\item
The Jost solution $f\left(  k,x\right)  $ has the representation
\begin{equation}
f\left(  k,x\right)  =e^{ikx}I+\int_{x}^{\infty}e^{iky}K\left(  x,y\right)
dy. \label{jostK}%
\end{equation}
\end{enumerate}
\end{prop}

The scattering matrix,
$S\left(  k\right),$  is a $n\times n$ matrix-valued function of $k\in \mathbb R$ that is given by
\begin{equation}
S\left(  k\right)  =-J\left(  -k\right)  J\left(  k\right)  ^{-1},\text{ }%
k\in\mathbb{R}\text{.} \label{Scatteringmatrix}%
\end{equation}

In the exceptional case where $J(0)$ is not invertible the scattering matrix is defined  by \eqref{Scatteringmatrix} only for $ k \neq 0.$ However, it is proven in \cite{[5]}, and in Theorem  3.8.14 in page 138 of \cite{WederBook}, that for potentials satisfying (\ref{PotentialHermitian}) and (\ref{PotentialL11})
the limit $S(0):=\lim_{k\to 0} S(k)$ exists in the exceptional case and, moreover,  a formula for $S(0)$ is given.

It is proven in  \cite{[9]}   and in Theorem 3.10.6 in pages  196-197 of \cite{WederBook}) that the following limit exist,
\begin{equation}
S_{\infty}:=\lim_{|k|\rightarrow\infty}S\left(  k\right).\label{sinft}%
\end{equation}

 Let us denote by $F_{s}$    the following  quantity,  that up to the  factor  $1 / \sqrt{2 \pi}$ is the inverse    Fourier transform
of    $S\left( k\right)  -S_{\infty},$

\begin{equation}  \label{fs}
F_{s}\left(  y\right)  =\frac{1}{2\pi}\int_{-\infty}^{\infty}\left[  S\left(
k\right)  -S_{\infty}\right]  e^{iky}dk,\text{ \ }y\in\mathbb{R}\text{.}
\end{equation}


 
The following theorem is proven in \cite{nawe}.

\begin{theorem}
\label{FourierL1}\textbf{ }Suppose that the potential $V$ satisfies
(\ref{PotentialHermitian}) and (\ref{PotentialL11}). Then,%
\begin{equation}
F_{s}\in L^{1}\left(  \mathbb{R }\right)  . \label{FourierSM}%
\end{equation}
\end{theorem}

In terms of the Jost solution $f(k,x)$ and the scattering matrix $S(k)$ we
construct the physical solution, \cite{[9]}  and equation (2.2.29) in page 26 of \cite{WederBook},%
\begin{equation}
\Psi\left(  k,x\right)  =f\left(  -k,x\right)  +f\left(  k,x\right)  S\left(
k\right)  ,\text{ }k\in\mathbb{R}\text{.} \label{Physicalsolution}%
\end{equation}
The physical solution $\Psi(k,x)$ is the main input  to construct the generalized Fourier
maps, $\mathbf F^\pm,$  for the absolutely continuous subspace of $H,$ that are defined in equation (4.3.44) in page 284 of \cite{WederBook} (see also Proposition 4.3.4 in page 287 of \cite{WederBook}),
\beq\label{gefoma}
\left(  \mathbf{F}^{\pm}Y\right)  \left(  k\right)  =\sqrt{\frac{1}{2\pi}%
}\int_{0}^{\infty}\left(  \Psi\left(  \mp k,x\right)  \right)  ^{\dagger}%
Y\left(  x\right)\, dx,
\ene 
for  $Y \in L^1(\mathbb R^+, \mathbb C^n) \cap L^2(\mathbb R^+, \mathbb C^n).$

We have ( see equation (4.3.46) in page 284 of\cite{WederBook}),
\begin{equation}
\label{isom}\left\Vert \mathbf{F}^{\pm}Y\right\Vert _{L^{2}(\mathbb R^+, \mathbb C^n)}= \left\Vert
E(\mathbb{R}^{+}; H)Y\right\Vert _{L^{2}(\mathbb R^+, \mathbb C^n)}.
\end{equation}
Thus, the $\mathbf{F}^{\pm}$ extend to bounded operators on $L^{2}(\mathbb R^+, \mathbb C^n)$ that we
also denote by $\mathbf{F}^{\pm}.$

The following results on the spectral theory  of $H$ are proven  in Theorem 3.11.1 in pages 199-200, Theorem 4.3.3 in page 284  and  Proposition 4.3.4 in page 287, of  \cite{WederBook}.

\begin{theorem}
\label{spect}Suppose that the potential $V$ satisfies
(\ref{PotentialHermitian}) and (\ref{PotentialL1}), and that the constant matrices $A,B$ fulfill \eqref{wcon1}, and \eqref{wcon2}. Then, the Hamiltonian $H$
has no positive eigenvalues and the negative spectrum of $H$ consists of
isolated eigenvalues of multiplicity smaller or equal than $n$, that can
accumulate only at zero. Furthermore, $H$ has no singular continuous spectrum
and its absolutely continuous spectrum is given by $[0,\infty)$. The
generalized Fourier maps $\mathbf{F}^{\pm}$ are partially isometric with
initial subspace $\mathcal{H}_{\operatorname*{ac}}\left(  H\right)  $ and
final subspace $L^{2}(\mathbb R^+,\mathbb C^n)$. Moreover, the adjoint operators are given by%
\beq \label{fad}
\left(  \left(  \mathbf{F}^{\pm}\right)  ^{\dagger}Z\right)  \left(
x\right)  =\sqrt{\frac{1}{2\pi}}\int_{0}^{\infty}\Psi\left(  \mp k,x\right)
Z\left(  k\right)  \,dk,
\ene
for $Z\in L^{1}(\mathbb R^+,\mathbb C^n) \cap L^{2}(\mathbb R^+,\mathbb C^n).$ Furthermore,
\begin{equation}
\mathbf{F}^{\pm}H\left(  \mathbf{F}^{\pm}\right)  ^{\dagger}=\mathcal{M},
\label{spectralrepr}%
\end{equation}
where $\mathcal{M}$ is the operator of multiplication by $k^{2}.$ If, in
addition, $V\in L_{1}^{1}(\mathbb R^+,M_n),$ zero is not an eigenvalue and the number of
eigenvalues of $H$ including multiplicities is finite.
\end{theorem}

Note that by (\ref{isom}) $\left(  \mathbf{F}^{\pm}\right)  ^{\dagger
}\mathbf{F}^{\pm}$ is the orthogonal projector onto $\mathcal{H}%
_{\operatorname*{ac}}\left(  H\right) ,$ 
\begin{equation}
\left(  \mathbf{F}^{\pm}\right)  ^{\dagger}\mathbf{F}^{\pm}%
=P_{\operatorname*{ac}}(H).\label{projector}%
\end{equation}

We denote by  $ F_0$  the cosine transform,
\beq\label{w.1}
\left(F_{0} Y \right)(k):= \sqrt{\frac{2}{\pi}}\, \int_0^\infty\,dx\, \cos(kx)\, Y(x), \quad Y \in L^2(\mathbf R^+).
\ene
Actually, $F_0$ coincides with the generalized Fourier maps for $H_0$ given by Theorem~ \ref{spect}.

The following theorem, proven in  Theorem 4.4.3 in page 297 of \cite{WederBook}, gives the stationary formulae for the wave operators. 
\begin{theorem} \label{theoa.5.4.3} Suppose that $ V$ satisfies \eqref{PotentialHermitian} and \eqref{PotentialL1}. Then, the wave operators $W_\pm(H,H_0)$ exist and are complete. Further, the following the stationary formulae hold,
\beq \label{w.3}
W_\pm=  \left({\mathbf F}^\pm\right)^\dagger\, F_{0}.
\ene
\end{theorem}
\subsection{$L^p$- boundedness of the wave operators in the half-line}
We prepare the following proposition.
\begin{prop}\label{hilbert}
Suppose that $Y \in L^2(\mathbb R, \mathbb C^n).$ Then,
\beq \label{w.4}
 \mathcal F^{-1} \left(\chi_{\mathbb R^+}(k) \,( \mathcal F Y)(k)\right)(x)=  \ds \frac{i}{2}\, (\mathcal H Y)(x)+ \ds\frac{1}{2}\, Y(x), \, x \in \mathbb R,
 \ene
 and
 \beq \label{w.5}
 \mathcal F \left(\chi_{\mathbb R^+}(k) \,( \mathcal F^{-1} Y)(k)\right)(x)= \ds \frac{-i}{2}\, (\mathcal H Y)(x)+ \ds \frac{1}{2} Y(x), x \in \mathbb R.
 \ene
  \end{prop}
\noindent{\it Proof:}
 The proof is an immediate consequence of  equations (3) and (4) in page 60 of \cite{gs}.
 
 \bull

\noindent {\it   Proof of Theorem ~ \ref{bounded}:} Let us first take $ Y \in C^\infty_0(\mathbb R^+).$ Note that $\mathcal E_{\text{\rm even}}Y \in C^\infty_0(\mathbb R).$ By \eqref{w.1}
\beq\label{w.6}
(F_0 Y)(k)= (\mathcal  F \mathcal E_{\text{\rm even}} Y)(k), \qquad k \in \mathbb R^+.
\ene
By \eqref{jostK}, \eqref{Physicalsolution}, \eqref{fad}, \eqref{w.3}, and \eqref{w.6} 
\beq\label{w.21} 
\left(W_\pm Y\right)(x):= \sum_{j=1}^6\, T_\pm^{(j)}(x),
\ene
 where 
 \beq\label{w.22}
 T_\pm^{(1)}(x):= \ds \frac{1}{\sqrt{2 \pi}}\, \int_{\mathbb R}\, e^{\pm ikx}\, \chi_{\mathbf R^+}(k)\, (\mathcal F \mathcal E_{\text{\rm even}}  Y)(k)\, dk,  
 \ene
 \beq\label{w.23}
 T_\pm^{(2)}(x):= \ds \frac{1}{\sqrt{2 \pi}}\, \int_{x}^\infty\, dz\, K(x,z)\, \int_{\mathbb R}\, e^{\pm i k z}\, \chi_{\mathbb R^+}(k)\, (\mathcal F \mathcal  E_{\text{\rm even}} Y)(k)\, dk,
 \ene
  \beq\label{w.24}
 T_\pm^{(3)}(x):= \ds \frac{1}{\sqrt{2 \pi}}\, \int_{\mathbb R}\, e^{\mp ikx}\, \chi_{\mathbb R^+}(k)\, S_\infty  (\mathcal F\mathcal  E_{\text{\rm even}}  Y)(k)\, dk,  
 \ene
  \beq\label{w.25}
 T_\pm^{(4)}(x):= \ds \frac{1}{\sqrt{2 \pi}}\, \int_{x}^\infty\, dz\, K(x,z)\, \int_{\mathbb R}\, e^{\mp i k z}\, \chi_{\mathbb R^+}(k)\, S_\infty  (\mathcal F \mathcal  E_{\text{\rm even}} Y)(k)\, dk,
 \ene
   \beq\label{w.26}
 T_\pm^{(5)}(x):= \ds \frac{1}{\sqrt{2 \pi}}\, \int_{\mathbb R}\, e^{\mp ikx}\, \chi_{\mathbf R^+}(k)\, (S(\mp k)-S_\infty)  (\mathcal F \mathcal  E_{\text{\rm even}} Y)(k)\, dk,  
 \ene
 and
  \beq\label{w.27}
  T_\pm^{(6)}(x):= \ds \frac{1}{\sqrt{2 \pi}}\, \int_{x}^\infty\, dz\, K(x,z)\, \int_{\mathbb R}\, e^{\mp i k z}\, \chi_{\mathbb R^+}(k)\,(S(\mp k)- S_\infty)  (\mathcal F \mathcal  E_{\text{\rm even}} Y)(k)\, dk.
\ene
Observe that
\beq\label{w.27.ddd}
(\mathcal F \mathcal E_{\rm even}Y)(k)= (\mathcal F^{-1} \mathcal E_{\rm even}Y)(k), \qquad k \in \mathbb R.
\ene
It    follows from Proposition ~\ref{hilbert}, and \eqref{w.21}-\eqref{w.27.ddd}  that \eqref{w.14} holds for $Y \in  C^\infty_0(\mathbb R^+).$ Finally,  approximating $ Y \in L^p(\mathbb R^+, \mathbb C^n),  1 < p < \infty, $  by a sequence  of functions in $C^\infty_0(\mathbb R^+),$ it follows  that equation \eqref{w.14}- \eqref{w.17}  hold for all $Y \in L^p(\mathbb R^+, \mathbb C^n), 1 < p < \infty,$ and that the wave operators $W_\pm(H, H_0)$ extend to bounded operators on $L^p(\mathbb R^+, \mathbb C^n), 1 < p < \infty.$  Here  we used that $ Q(F_{\rm s})$ is bounded in $L^p(\mathbb R, \mathbb C^n), 1 \leq p \leq \infty,$ since by Theorem~\ref{FourierL1} $F_s \in L^1(\mathbb R, \mathbb C^n),$  that ${\mathbf K}(K)$  is bounded in $L^p(\mathbb R^+, \mathbb C^n), 1 \leq p \leq \infty,$ because by \eqref{ESTK}, equations \eqref{w.13.c} hold, and that $\mathcal H$ is bounded in $L^p(\mathbb R, \mathbb C^n), 1  < p < \infty$ \cite{sado}, \cite{stein}. 
 The operator $\mathcal R$ is clearly bounded from $L^p(\mathbb R, \mathbb C^n)$ into $L^p(\mathbb R^+, \mathbb C^n), 1 \leq p \leq \infty.$
 The wave operator $W_\pm(H,H_0)^\dagger$ extend to bounded operators on $L^p(\mathbb R^+, \mathbb C^n), 1 < p < \infty,$  by duality.
 
 \qed

   \noindent {\it   Proof of Theorem~\ref{theo.new.1}:}
Let us first take $ Y \in C^\infty_0(\mathbb R^+, \mathbb C^n).$ Recall  that $\mathcal E_{\text{\rm even}}Y \in C^\infty_0(\mathbb R),$ and that  \eqref{w.6} holds.
By \eqref{w.22}, \eqref{w.24}, $S_\infty=I_n$, and since $\mathcal F \mathcal E_{\rm even} Y$ is an even function,
\begin{align}\label{new.1}
&T^{(1)}_\pm(x)+ T^{(3)}_\pm(x)=  \ds \frac{1}{\sqrt{2 \pi}}\, \int_{\mathbb R}\, e^{\pm ikx}\, \chi_{\mathbf R^+}(k)\, (\mathcal F \mathcal E_{\text{\rm even}}  Y)(k)\, dk+\\ \nonumber
& \ds \frac{1}{\sqrt{2 \pi}}\, \int_{\mathbb R}\, e^{\mp ikx}\, \chi_{\mathbb R^+}(k)\,  (\mathcal F\mathcal  E_{\text{\rm even}}  Y)(k)\, dk= 
 \mathcal E_{\rm even}Y(x)= Y(x), \qquad x \geq 0,
 \end{align}   
 where in the second integral in the middle equation in \eqref{new.1} we made the change of variable of integration $ k \to -k,$ and we used, $\mathcal F \mathcal E_{\rm even}
 Y(k)=  \mathcal F^{-1} \mathcal E_{\rm even}Y(k),$ and $S_\infty=I_n.$
We similarly prove, using\eqref{w.23} and \eqref{w.25},
\beq\label{new.2}
T^{(2)}_\pm(x)+ T^{(4)}_\pm(x)= \left(\mathbf K(K) Y\right)(x).
\ene
Hence, by \eqref{w.21}, \eqref{new.1}, and \eqref{new.2},
\beq\label{new.3}
\left( W_\pm Y\right)(x)= Y(x)+ \ds \left(\mathbf K(K)Y\right)(x)+ T^{(5)}_\pm(x)+T^{(6)}_\pm(x).
\ene
By  \eqref{w.26}, \eqref{w.27} and the convolution theorem of the Fourier transform,
\beq\label{new.4}
T^{(5)}_\pm(x)=( Q(P_\pm)\mathcal E_{\rm even}Y)(x), \qquad x \geq 0,
\ene
and,
\beq\label{new.5}
T^{(6)}_\pm(x)= \left(\mathbf K(K)\mathcal R\, Q(P_\pm)\mathcal E_{\rm even}Y\right)(x),
\ene
where $P_\pm$ is defined in \eqref{new.0000}.
Equation \eqref{new.000.aa}  for $ Y \in C^\infty_0(\mathbb R^+, \mathbb C^n)$ follows from \eqref{new.3}, \eqref{new.4} and \eqref{new.5}. Moreover, as  $\mathcal R$ is bounded from $L^p(\mathbb R, \mathbb C^n)$ into $L^p(\mathbb R^+,\mathbb C^n), 1 \leq p \leq \infty,$ $\mathcal E_{\rm even}$ is bounded from $L^p(\mathbb R^+, \mathbb C^n)$ into $L^p(\mathbb R,\mathbb C^n), 1 \leq p\leq \infty,$   $\mathbf K(K)$  is bounded in $L^p(\mathbb R^+, \mathbb C^n), 1 \leq p \leq \infty,$  because by \eqref{ESTK} equations\eqref{w.13.c}  hold.  Moreover, by Schwarz inequality,
  \beq\label{new.5.b}\begin{array}{l}
  \| P_\pm\|_{L^1(\mathbb R, M^n)}\leq \| (1+|x|^2)^{-1/2}\| _{L^2(\mathbb R^+)}\, \| (1+|x|^2)^{1/2} P_\pm \|_{L^2(\mathbb R, M^n)}\leq\\ \\
  C\left\| \chi_{\mathbf R^+}(k)\, (S( \mp k)-S_{\infty})\right\|_{\mathbf H^{(1)}(\mathbb R, M_n)}.
  \end{array}
  \ene
  By the definition of $S(k)$ in \eqref{Scatteringmatrix} and by Proposition 3.2.4 in page 65, Theorem 3.81 in page 114 and Theorem 3.9.15 in pages 189-190, of \cite{WederBook} $S(k)$ is differentiable for $k \in \mathbb R,$ with continuous derivative for $k \in \mathbb R\setminus \{0\}.$  Then, since $S(0)=S_\infty,$   and as by Proposition~\ref{prop.h1} $S(k)-S_\infty \in  \mathbf H^{(1)}(\mathbb R^+, M_n)$ we have that
 $ \chi_{\mathbf R^+}(k)\, (S(\mp k)-S_{\infty})\in  \mathbf H^{(1)}(\mathbb R^+, M_n).
 $
 Hence, by \eqref{new.5.b}, 
 $
 P_\pm\in L^1(\mathbb R^+, M^n),
 $
 and then, $Q(P_\pm)  $ is a bounded operator in $L^p(\mathbb R, \mathbb C^n), 1 \leq p \leq \infty.$ Hence, \eqref{new.000.aa} holds for all $ Y \in L^2(\mathbb R^+, \mathbb C^n)$ and, moreover,  the wave operators $W_\pm(H,H_0)$ extend to bounded operators in $L^1(\mathbb R^+,\mathbb C^n)$ and in  $L^\infty(\mathbb R^+,\mathbb C^n).$

 
 We now prove that the adjoint wave operators $W_\pm(H,H_0)^\dagger$ extend to bounded operators in $L^1(\mathbb R^+,\mathbb C^n)$ and in  $L^\infty(\mathbb R^+,\mathbb C^n).$ By \eqref{new.000.aa}, 
 \beq\label{new.5.d}
\left( W_\pm^\dagger Y\right)(x)= Y(x)+ \ds \left(\mathbf K^\dagger(K)Y\right)(x)+\left(\mathcal E^\dagger_{\rm even} Q(P_\pm)^\dagger \mathcal R^\dagger Y\right)(x)+\left(\mathcal E^\dagger_{\rm even}  Q(P_\pm)^\dagger   \mathcal R^\dagger \mathbf K(K)^\dagger Y\right)(x).
\ene
We have,
$$
\mathbf K(K)^\dagger Y(x)= \int_0^x\, K^\dagger(y,x)\, Y(y)\, dy.
$$
By \eqref{ESTK}, equation \eqref{w.13.c}  holds, and then $\mathbf K(K)^\dagger$ is bounded in $L^1(\mathbb R^+, \mathbb C^n)$ and  in $L^\infty(\mathbb R^+,\mathbb C^n).$ Further,
$$
Q_{}^\dagger (P_\pm) Y(x)= \int_{-\infty}^\infty\, P^\dagger_\pm(y-x)\, Y(y)\, dy,
$$
and as $P^\dagger_\pm \in   L^1(\mathbb R, \mathbb C^n),$ it follows that $Q^\dagger(P_\pm)$ in bounded $L^1(\mathbb R, \mathbb C^n)$ and  in $L^\infty(\mathbb R,\mathbb C^n).$ Moreover,
$
\mathcal E_{\rm even}^\dagger Y(x)= Y(x)+Y(-x),
$
 and then, $\mathcal E_{\rm even}^\dagger $  is bounded from  $L^1(\mathbb R, \mathbb C^n)$  into $L^1(\mathbb R^+, \mathbb C^n)$ and from  $L^\infty(\mathbb R, \mathbb C^n)$  into $L^\infty(\mathbb R^+, \mathbb C^n).$ Furthermore,
 $$
 \mathcal R^\dagger Y(x)= \left\{\begin{array}{l} Y(x),  x \geq 0, \\ 0, \qquad x < 0,
 \end{array}\right.
 $$
 and it follows that  $\mathcal R^\dagger $ is bounded from  $L^1(\mathbb R, \mathbb C^n)$  into $L^1(\mathbb R^+, \mathbb C^n)$ and from  $L^\infty(\mathbb R, \mathbb C^n)$  into \linebreak $L^\infty(\mathbb R^+, \mathbb C^n).$  Then,  by \eqref{new.5.d} the adjoint wave operators $W_\pm(H,H_0)^\dagger$  extend to bounded operators in $L^1(\mathbb R^+,\mathbb C^n)$ and in  $L^\infty(\mathbb R^+,\mathbb C^n).$ 

  \qed
  \begin{remark}\label{rem.new2}
  {\rm The condition $S_\infty=S(0)=I_n$ is actually necessary in Theorem~\ref{theo.new.1}, as the following example shows. Consider the scalar case, $n=1,$ with $V=0,$  Dirichlet boundary condition, $Y(0)=0,$  and boundary matrices, $B=-1, A=0.$ In this case by equation (3.7.5) in page 113 of \cite{WederBook}, $S_\infty=S(0)=- 1.$
  Further by equation (4.3.8) in page 277 of \cite{WederBook} the generalized Fourier maps are given by,
  \beq\label{new.6}
  (\mathbf F^\pm Y)(k):=\mp 2  i \frac{1}{\sqrt{2\pi}}\, \int_0^\infty\, \sin kx\, Y(x)\,dx.
  \ene
  Hence, by \eqref{w.3} and \eqref{w.6}
  \beq\label{new.7}
  W_\pm(H,H_0) Y=\pm \mathcal R  \mathcal F^{-1}\, {\rm sign}\, k \mathcal F\mathcal E_{\rm even} Y.
  \ene
 Moreover,we have
 \beq\label{new.8}
 \mathcal F^{-1} \chi_{\mathbb R^-}(k)\, \mathcal F \mathcal E_{\rm even}Y= \mathcal F \chi_{\mathbb R_+}(k)\mathcal F^{-1}\mathcal E_{\rm even} Y.
 \ene
 Then, by \eqref{w.4}, \eqref{w.5}, \eqref{new.7}, \eqref{new.8},
 \beq\label{new.9}
 W_\pm(H,H_0) Y= \pm i   \mathcal R \mathcal H \mathcal E_{\rm even}Y.
 \ene
 Finally, since the Hilbert transform is not bounded in $L^1(\mathbb R, \mathbb C)$ and in $L^\infty(\mathbb R,\mathbb C)$ \cite{sado}, \cite{stein}, it follows that $W_\pm(H,H_0)$ are not bounded
 $L^1(\mathbb R^+, \mathbb C)$ and in $L^\infty(\mathbb R^+,\mathbb C).$}
 \end{remark}

\subsection{The $L^p$- boundedness of the wave operators on the line} 
 \noindent {\it Proof of Theorem ~\ref{wholeline}:}
 We first prepare some results, Let us denote by $H_{1}$ the Hamiltonian  $H_{A,B,V}$ with the matrices given in \eqref{matrices2} with $\Lambda=0,$  and with the potential $V$ identically zero. Note that
 
 \beq\label{w.28}
 H_{0,\mathbb R}= \mathbf U \, H_1\, \mathbf U^\dagger.
 \ene
 
 By Theorem~\ref{theoa.5.4.3} the wave operators
 $
 W_\pm(H_{1}, H_0),
 $
exist and are complete. Then,  by Proposition 3 in page 18 of \cite{rs3} the wave operators
$
W_\pm(H_0, H_{1}),
$
also exists and are complete,
and furthermore,
\beq\label{w.29}
W_\pm(H_{0}, H_{1})= W_\pm( H_{1}, H_0)^\dagger.
\ene
Then, by Theorem ~\ref{bounded}  the wave operators $W_\pm(H_{0}, H_{1}),$   restricted to  $L^2(\mathbb R^+, \mathbb C^{2n})\cap L^p(\mathbb R^+, \mathbb C^{2n}), 1 <  p < \infty,$ extend uniquely to bounded  operators in $L^p(\mathbb R^+, \mathbb C^{2n}), 1 < p < \infty.$ Further, by the chain rule, see  Proposition 2 in page 18 of \cite{rs3},
\beq\label{w.30}
W_\pm(H,H_{1})= W_\pm(H, H_0)\, W_\pm(H_0, H_{1}).
\ene
Hence,  as   $ W_\pm(H, H_0),$ and  $W_\pm(H_0, H_{1}),$ restricted to  $L^2(\mathbb R^+, \mathbb C^{2n})\cap L^p(\mathbb R^+, \mathbb C^{2n}), 1 <  p < \infty,$ extend uniquely to bounded  operators in $L^p(\mathbb R^+,\mathbb C^{2n}), 1 < p < \infty,$ if follows that also $W_\pm(H, H_{1}), $restricted to  $L^2(\mathbb R^+, \mathbb C^{2n})\cap L^p(\mathbb R^+, \mathbb C^{2n}), 1 <  p < \infty,$  extend uniquely to bounded  operators in $L^p(\mathbb R^+, \mathbb C^{2n}), 1 < p < \infty.$   Finally,  by \eqref{hwhole} and \eqref{w.28},
 \beq\label{w.31}
 W_\pm(H_\mathbb R, H_{0,\mathbb R})= \mathbf U\, W_\pm(H,H_{1})\, \mathbf U^\dagger,
 \ene 
 and, as $\mathbf U$ is bounded from $L^p(\mathbb R^+, \mathbb C^{2n}),$  into $L^p(\mathbb R, \mathbb C^{n}),$  and $\mathbf U^\dagger$ is bounded from  $L^p(\mathbb R, \mathbb C^n),$ into $L^p(\mathbb R^+, \mathbb C^{2n}), 1 \leq p \leq \infty,$ we have that the 
  $W_\pm(H_\mathbb R, H_{0,\mathbb R})$ restricted to    $L^2(\mathbb R, \mathbb C^n)\cap L^p(\mathbb R, \mathbb C^n), 1 <  p < \infty,$ extend uniquely to bounded  operators in $L^p(\mathbb R, \mathbb C^n), 1 < p < \infty.$ Finally, by duality the adjoint wave operators $W_\pm(H_\mathbb R, H_{0,\mathbb R})^\dagger$  extend uniquely to bounded  operators in $L^p(\mathbb R, \mathbb C^n), 1 < p < \infty.$ 
  
  \bull

  We now proceed to  prove that the wave operators $W_\pm(H_{\mathbb R}, H_{0,\mathbb R})$ are bounded in $L^1(\mathbb R, \mathbb C^n),$ and in $L^\infty(\mathbb R, \mathbb C^n),$ as stated in Theorem ~\ref{theo.new.6}. We first prepare some results. 
  
  Let us denote by $A_1,B_1$ the matrices \eqref{matrices2} with $\Lambda=0.$ Then,
  $
  H_1= H_{A_1, B_1, 0}.
  $ 
 Let $ \tilde{A}_1, \tilde{B}_1$ be the matrices related to $A_1, B_1$  as in \eqref{A,Btilde}, \eqref{PSM13}, and \eqref{trans} for some invertible matrices $T_{1,1}, T_{2,1}$ and some unitary matrix $ \mathcal M_1.$ Hence,
 $$
 A_1= \mathcal M_1 \tilde{A}_1 T_{1,1} \mathcal M_1^\dagger T_{2,1}, \qquad B_1= \mathcal M_1\tilde{B}_1 T_{1,1} \mathcal M_1^\dagger T_{2,1}.
 $$
 To simplify the notation we denote,
 $
 \tilde{H}_1:= H_{\tilde{A}_1, \tilde{B}_1,0}.
 $
 Applying \eqref{diagonalization} to $H_1$ and $\tilde{H}_1$ we obtain,
 \beq\label{new.10}
 H_1= \mathcal M_1 \tilde{H}_1\mathcal M_1^\dagger.
 \ene
 Let us denote by $\mathbf F^\pm_1,$ respectively $\tilde{\mathbf F}^\pm_1,$  the generalized Fourier maps for $H_1,$ and for $\tilde{H}_1$ defined in \eqref{gefoma}. Then, by   \eqref{new.10} we get (see equation (4.3.35)  in page 282  of \cite{WederBook})  
 \beq\label{new.11}
 \mathbf F^\pm_1= \mathcal M_1 \tilde{\mathbf F}^\pm_1 \mathcal M_1^\dagger.
 \ene
 By \eqref{w.3}, \eqref{w.29}, \eqref{w.30} and as 
 $F_0=F_0^\dagger= F_0^{-1},$
 $
 W_\pm(H, H_1)= \left(\mathbf F^\pm\right)^\dagger\,\mathbf  F^\pm_1,
 $ 
 and using \eqref{new.11} we prove,
 \beq\label{new.12}
W_\pm(H, H_1)=\left( \mathbf F^\pm\right)^\dagger\, \mathcal M_1  \tilde{\mathbf F}^\pm_1 \mathcal M_1^\dagger.
\ene
To use \eqref{new.12} to study the boundedness of the wave operators we need to compute explicitly the unitary matrix $\mathcal M_1.$ 
For this purpose, we first introduce the following unit vectors in $\mathbb C^{2n},$
\beq\label{new.13}   
Y^{(j)}:= (0,\dots,\frac{1}{\sqrt 2},0,0\dots,- \frac{1}{\sqrt 2},0,\dots,0 )^T, \quad j=1,1,\dots, n,
\ene
with components that take the  value $\frac{1}{\sqrt{ 2}}$ at the component $j,$ the value $-\frac{1}{\sqrt {2}}$ at the component  $j+n, 1\leq j \leq n,$ and all the other components are zero.
Further, we define the following unit vectors in $\mathbb C^{2n}$ 
\beq\label{new.14}   
Y^{(j)}:=    (0,\dots,\frac{1}{\sqrt 2},0,0\dots, \frac{1}{\sqrt 2},0,\dots,0 )^T, \quad j=n+1, \dots, 2n,
\ene
with components that take the  value $\frac{1}{\sqrt 2}$ at the component $j-n,$ the value $\frac{1}{\sqrt 2}$ at the component  $j,$ $n+1\leq j \leq 2n,$ and all the other components are zero.

\begin{prop}\label{prop.new3}
Let $A_1,B_1$ be the matrices  \eqref{matrices2} with $\Lambda=0,$ and let  $ \tilde{A}_1, \tilde{B}_1$ be the matrices related to $A_1, B_1$  as in \eqref{A,Btilde}, \eqref{PSM13}, and \eqref{trans} for some invertible matrices $T_{1,1}, T_{2,1}$ and some unitary matrix $\mathcal M_1.$ Then,
\begin{enumerate}
\item 
\beq\label{new.15}
\tilde{A}_1=\left\{
\begin{array}
[c]{lc}%
0_{n} &  0_{n}\\
0_{n} & -I_{n}%
\end{array}
\right\}. 
\ene

\item
\beq\label{new.16}
\tilde{B}_1=\left\{
\begin{array}
[c]{lc}%
-I_{n} &  0_{n}\\
0_{n} & 0_{n}%
\end{array}
\right\}. 
\ene
\item
The boundary conditions \eqref{PSM13} are given by,
\beq\label{new.17}
Y_j(0)=0,\,\, j=1,\dots,n,\qquad Y'_j(0)=0, \,\ j= n+1,\dots, 2n.
\ene
That is to say, the first $n$ components of $Y$ satisfy the Dirichlet boundary condition, and the last $n$ components fulfill the Neumann boundary condition.
\item
The unitary matrix, $\mathcal M_1$, is given by,
\beq\label{new.18}
\mathcal M_1= \left\{ Y^{(1)}\, Y^{2} \dots Y^{(2n)}  \right\}.
\ene
\item
The invertible matrices $T_{1,1},$ and $T_{2,1}$ are given by,
\beq\label{new.19}
T_{1,1}=\left\{
\begin{array}
[c]{lc}%
-I_{n} &  0_{n}\\
0_{n} & i I_{n}%
\end{array}\right\},
\ene 

\beq\label{new.20}
T_{2,1}=\left\{
\begin{array}
[c]{lc}%
-I_{n} & i I_{n}\\
I_{n} & iI_{n}%
\end{array}
\right\}. 
 \ene
\end{enumerate}
\end{prop}

\noindent {\it Proof:} We use the notation of the proof of Proposition 3.4.5 in pages 101-103 of \cite{WederBook}.   We denote,
$
E:=\sqrt{ A_1^\dagger A_1+ B_1^\dagger B_1},
$
 and
 $
 U:= (B_1-iA_1) E^{-2} (B_1^\dagger- i A_1^\dagger). 
 $
Then, by \eqref{matrices2} and a simple computation, we get,
$$
U=\left\{ \begin{matrix} 0& -I_n \\-I_n & 0
\end{matrix}\right\}.
$$
 It is easily verified that the $Y^{(j)}, j=1,\dots,n$ are eigenvectors of $U$ with eigenvalues  one, and that the vectors  $ Y^{(j)}, j=n+1,\dots, 2n$ are  eigenvectors of $U$
 with eigenvalue minus one. Then, the columns  of $\mathcal M_1$ are an orthonormal system of eigenvectos of $U,$ and in consequence $\mathcal M_1$ diagonalizes $U,$ as required in equation (3.4.39) in page 101 of \cite{WederBook},
 $
 \mathcal M_1^\dagger U \mathcal M_1= {\rm diag}\{1,\dots,1,-1,\dots,-1\},
 $
 is the matrix with the first $n$ diagonal entries equal to  one, the second $n$ diagonal entries equal to minus one and all other entries equal to zero. This proves that item (4) is satisfied.
 Using the notation of equation (3.4.41) in page 102 of \cite{WederBook} with $P=I_{2n}$ we get
 $
 \mathcal M_1^\dagger U \mathcal M_1= {\rm diag}\{ e^{2\theta_1},\dots, e^{2i\theta_{2n}} \}= {\rm diag}\{1,\dots,1,-1,\dots,-1\}, 0< \theta_j \leq \pi, j=1,\dots, 2n.
 $
 Then, 
 $ 
 \theta_1=\theta_2=\dots=\theta_{n}=\pi, \, {\rm and}\, \theta_{n+1}= \theta_{n+2}=\dots=\theta_{2n}=\pi/2,
 $
 and items (1), (2) and (3) hold. That item (5) holds is immediate since by the definition of $T_{1,1}$ and $T_{2,1}$ in page 103 of \cite{WederBook}, we have
 $T_{1,1}:= (\tilde{B}_1+i \tilde{A}_1)^{-1},$ and $T_{2,1}:= B_1+iA_1.$
 
 \qed

 Let us denote  $ \tilde{\psi}^+_1(k,x):= \tilde{\psi}_1(-k,x),$ and  $ \tilde{\psi}^-_1(k,x):= \tilde{\psi}_1(k,x),$ where $\tilde{\psi}_1(k,x)$ is  the physical solution of $\tilde{H}_1.$
 By equations (4.3.6) and  (4.3.7) in page 277 of \cite{WederBook},
 \beq\label{new.21.b}
 \tilde{\psi}^\pm_1(k,x)=\{\pm 2 i \sin kx,\dots,\pm 2i \sin kx, 2\cos kx,\dots, 2 \cos kx\}, 
 \ene
 is the diagonal matrix  with the first $n$ diagonal components equal to $\pm 2 i \sin kx,$ and the last $n$ diagonal components equal to $2 \cos kx.$ Further, by equation (4.3.8) in page 277 of \cite{WederBook},
 \beq\label{new.22}
 \left(  \tilde{\mathbf F}^\pm_1 Y\right)(k)= \sqrt{\frac{1}{2\pi}}\, \int_0^\infty\, \tilde{\psi}^\pm_1(k,x)^\dagger\, Y(x)\, dx.
 \ene
 By \eqref{new.21.b} and \eqref{new.22},
 \beq\label{new.23}
 \left( \tilde{\mathbf F}^\pm_1 Y\right)(k)=\mathcal R  \left(\pm  (\mathcal F \mathcal E_{\rm odd} Y_+)(k), (\mathcal F\mathcal E_{\rm even} Y_-)(k) \right)^T.
 \ene
 We denote,
 \beq\label{new.24}
 W_{\pm,\mathcal M_1}(H,H_1):= \mathcal M_1^\dagger W_\pm(H, H_1) \mathcal M_1,
 \ene
 \beq\label{new.25}
 S_{\mathcal M_1}(k):= \mathcal M_1^\dagger S(k) \mathcal M_1,\qquad k \in \mathbb R.
 \ene 
 and
  \beq\label{new.26}
 S_{\infty, \mathcal M_1}:= \mathcal M_1^\dagger S_\infty \mathcal M_1,\qquad k \in \mathbb R.
 \ene 
Using \eqref{Physicalsolution}, \eqref{fad}, \eqref{new.12}, and  \eqref{new.23}-\eqref{new.26} we  prove,
\beq\label{new.27} 
\left(W_{\pm ,\mathcal M_1} Y\right)(x):= \sum_{j=1}^6\, J_\pm^{(j)}(x),
\ene
 where 
 \beq\label{new.28}
 J_\pm^{(1)}(x):= \ds \frac{1}{\sqrt{2 \pi}}\, \int_{\mathbb R}\, e^{\pm ikx}\, \chi_{\mathbf R^+}(k)\,  \left(\pm  (\mathcal F \mathcal E_{\rm odd} Y_+)(k), (\mathcal F\mathcal E_{\rm even} Y_-)(k) \right)^T\, dk,  
 \ene
 \beq\label{new.29}
 J_\pm^{(2)}(x):= \ds \frac{1}{\sqrt{2 \pi}}\, \mathcal M^\dagger_1\int_{x}^\infty\, dz\, K(x,z)\,\mathcal M_1\, \int_{\mathbb R}\, e^{\pm i k z}\, \chi_{\mathbb R^+}(k)\,  \left(\pm  (\mathcal F \mathcal E_{\rm odd} Y_+)(k), (\mathcal F\mathcal E_{\rm even} Y_-)(k) \right)^T\,dk.
 \ene
  \beq\label{new.30}
 J_\pm^{(3)}(x):= \ds \frac{1}{\sqrt{2 \pi}}\, \int_{\mathbb R}\, e^{\mp ikx}\, \chi_{\mathbb R^+}(k)\, S_{\infty, \mathcal M_1}   \left(\pm  (\mathcal F \mathcal E_{\rm odd} Y_+)(k), (\mathcal F\mathcal E_{\rm even} Y_-)(k) \right)^T\, dk,  
 \ene
  \beq\label{new.31}
 J_\pm^{(4)}(x):= \ds \frac{1}{\sqrt{2 \pi}}\, \int_{x}^\infty\, dz\, \mathcal M^\dagger_1 K(x,z) \mathcal M_1\, \int_{\mathbb R}\, e^{\mp i k z}\, \chi_{\mathbb R^+}(k)\, S_{\infty, \mathcal M_1}  \left(\pm  (\mathcal F \mathcal E_{\rm odd} Y_+)(k), (\mathcal F\mathcal E_{\rm even} Y_-)(k) \right)^T \, dk,
 \ene
   \beq\label{new.32}
 J_\pm^{(5)}(x):= \ds \frac{1}{\sqrt{2 \pi}}\, \int_{\mathbb R}\, e^{\mp ikx}\, \chi_{\mathbf R^+}(k)\, (S_{\mathcal M_1}(\mp k)-S_{\infty,\mathcal M_1})  \left(\pm  (\mathcal F \mathcal E_{\rm odd} Y_+)(k), (\mathcal F\mathcal E_{\rm even} Y_-)(k) \right)^T \, dk,  
 \ene
 and
  \begin{align}\label{new.33}\nonumber
  J_\pm^{(6)}(x)&:= \ds \int_{x}^\infty\, dz\,\mathcal M^\dagger_1 K(x,z) \mathcal M_1\,\frac{1}{\sqrt{2 \pi}}\,  \int_{\mathbb R}\, e^{\mp i k z}\, \chi_{\mathbb R^+}(k)\,(S_{\mathcal M_1}(\mp k)- S_{\infty,\mathcal M_1})  \cdot
  \\&\left(\pm  (\mathcal F \mathcal E_{\rm odd} Y_+)(k), (\mathcal F\mathcal E_{\rm even} Y_-)(k) \right)^T dk.
\end{align}
We denote,
\beq\label{new.34}
P_{\pm, \mathcal M_1}(x):= \mathcal M^\dagger_1 P_\pm(x) \mathcal M_1,
\ene
where $P_\pm$ is defined in \eqref{new.0000}.
   \begin{theorem} \label{theo.new.4} 
  Suppose that $V$ fulfills \eqref{PotentialHermitian}, that $V \in L^1_\gamma(\mathbb R^+,M_{2n}), \gamma > \frac{5}{2}$, that the constant  matrices $A,B$ are given by \eqref{matrices2} with $\Lambda=0,$ and that  
  \beq\label{new.35}
  S(0)= S_{\infty}= \left\{
\begin{array}
[c]{lc}%
0_n &  I_n\\
I_n& 0_n%
\end{array}
\right\},
\ene  
 where, $0_n,$ and $I_n$ are, respectively,  the  $n\times n$  zero matrix and the $n \times n$ identity matrix.
  
  Then, for all $Y \in L^2(\mathbb R^+, \mathbb C^{2n})$ we have,
  \begin{align}\label{new.36}
W_{\pm}(H,H_1)Y& = Y+\mathbf K(K) Y+ \mathcal M_1 \mathcal R Q(P_{\pm, \mathcal M_1})  \left(-  \mathcal E_{\rm odd} (\mathcal M^\dagger_1Y)_+, \mathcal E_{\rm even}(\mathcal M^\dagger _1Y)_-\right)^T \\\nonumber
+& \mathbf K(K) \mathcal M_1 \mathcal R\, Q( P_{\pm, \mathcal M_1})  \left(-  \mathcal E_{\rm odd} (\mathcal M^\dagger_1Y)_+, \mathcal E_{\rm even}(\mathcal M^\dagger _1Y)_- \right)^T.
\end{align}
Furthermore, the wave operators $W_\pm(H,H_1)$ and $W_\pm(H,H_1)^\dagger$ restricted to $L^2(\mathbb R^+,\mathbb C^{2n})\cap L^1(\mathbb R^+,\mathbb C^{2n})$, respectively to $L^2(\mathbb R^+,\mathbb C^{2n})\cap L^\infty(\mathbb R^+,\mathbb C^{2n}),$ extend to bounded operators on $L^1(\mathbb R^+,\mathbb C^{2n})$ and to bounded operators on $L^\infty(\mathbb R^+,\mathbb C^{2n}).$
 The $2n\times2 n$ matrix valued function $K(x,y), x, y \in \mathbb R^+$ is defined in  \eqref{marchenko}. Moreover, the $2n \times 2n$ matrix valued function $P_{\pm, \mathcal M_1}(x)$ is defined in \eqref{new.34}. The scattering matrix $S(k), k \in \mathbb R,$ is defined in \eqref{Scatteringmatrix} and the quantity $S_{\infty}$ in \eqref{sinft}.
\end{theorem}
  \noindent {\it   Proof of Theorem~\ref{theo.new.4} :}
  By \eqref{new.18}, \eqref{new.25},  \eqref{new.26}, \eqref{new.35},

   \beq\label{new.37}
  S_{\mathcal M_1}(0)= S_{\infty,\mathcal M_1}= \left\{ \begin{array}{lc} -I_n& 0_n\\ 0_n& I_n\end{array} \right\}.
  \ene
  Then, by \eqref{new.27}-\eqref{new.34}, and \eqref{new.37},
  \begin{align}\label{new.38}
&W_{\pm,\mathcal M_1}(H,H_1)Y = Y+\mathcal M^\dagger_1\mathbf K(K) \mathcal M_1 Y+  \mathcal R Q(P_{\pm, \mathcal M_1})  \left(- \mathcal E_{\rm odd} Y_+, \mathcal E_{\rm even}Y_-\right)^T \\ \nonumber
&+\mathcal M^\dagger_1 \mathbf K(K) \mathcal M_1 \mathcal R\, Q(P_{\pm, \mathcal M_1})  \left(-  \mathcal E_{\rm odd} Y _+, \mathcal E_{\rm even}(Y_- \right)^T.
\end{align}
  Equation \eqref{new.36} follows from \eqref{new.24} and \eqref{new.38}.  By \eqref{new.34},
  \beq\label{new.39}
  \| P_{\pm, \mathcal M_1}\|_{L^1(\mathbb R, M_{2n})} =   \| P_{\pm }\|_{L^1(\mathbb R,  M_{2n})}. 
  \ene
  As we already proved in the proof of Theorem~\ref{theo.new.1} that $ P_\pm \in L^1(\mathbb R,  M_{2n})$ it follows from \eqref{new.39} that $ P_{\pm, \mathcal M_1} \in L^1(\mathbb R,  M_{2n}).$ Hence, $Q(P_{\pm,\mathcal M_1})$ is bounded in   $L^1(\mathbb R,\mathbb C^{2n})$ and in $L^\infty(\mathbb R, \mathbb C^{2n}).$  We already proved in the proof Theorem~\ref{theo.new.1}  $\mathbf K(K),$ is  bounded in $L^1(\mathbb R^+, \mathbb C^{2n})$ and  in $L^\infty(\mathbb R^+,\mathbb C^{2n}),$  that $\mathcal R$ is bounded from  $L^1(\mathbb R, \mathbb C^{2n})$ into  $L^1(\mathbb R^+, \mathbb C^{2n})$ and from  $L^\infty(\mathbb R, \mathbb C^{2n})$ into  $L^\infty(\mathbb R^+, \mathbb C^{2n}).$  Clearly,  $\mathcal E_{\rm even}$ is bounded from  $L^1(\mathbb R,^+ \mathbb C^{2n})$ into  $L^1(\mathbb R, \mathbb C^{2n})$ and from  $L^\infty(\mathbb R^+, \mathbb C^{2n})$ into  $L^\infty(\mathbb R, \mathbb C^{2n}).$ Moreover, it is clear that   $\mathcal E_{\rm odd}$ is also bounded from  $L^1(\mathbb R,^+ \mathbb C^{2n})$ into  $L^1(\mathbb R, \mathbb C^{2n})$ and from  $L^\infty(\mathbb R^+, \mathbb C^{2n})$ into  $L^\infty(\mathbb R, \mathbb C^{2n}).$ Then, by \eqref{new.36} the wave operators $W_{\pm}(H,H_1)$ extend to bounded operators on $L^1(\mathbb R^+,\mathbb C^{2n})$ and to bounded operators on $L^\infty(\mathbb R^+,\mathbb C^{2n}).$  By  \eqref{new.38} and taking adjoints   we obtain,
  \beq\label{new.39.a} \begin{array}{l}
  W_{\pm, \mathcal M_1}^\dagger Y= Y+\mathcal  M^\dagger_1  \mathbf K(K)^\dagger \mathcal M_1 Y+\left( - \mathcal E^\dagger_{\rm odd}(Q(P_{\pm, \mathcal M_1})^\dagger \mathcal R^\dagger Y)_+, 
  \mathcal E_{\rm even}^\dagger(Q(P_{\pm, \mathcal M_1})^\dagger \mathcal R^\dagger Y)_- \right)+ \\
  \left(   - \mathcal E_{\rm odd}^\dagger  ( Q(P_{\pm, \mathcal M_1})^\dagger   \mathcal R^\dagger  \mathcal M_1^\dagger    \mathbf K(K)^\dagger \mathcal M_1 Y)_+,   \mathcal E^\dagger_{\rm even} ( Q(P_{\pm, \mathcal M_1})^\dagger   \mathcal R^\dagger  \mathcal M_1^\dagger    \mathbf K(K)^\dagger \mathcal M_1 Y)_- \right).
  \end{array}
  \ene
We already proved in the proof of Theorem~\eqref{theo.new.1} that $\mathbf K(K)^\dagger$ is bounded in $L^1(\mathbb R^+, \mathbb C^{2n})$ and  in $L^\infty(\mathbb R^+,\mathbb C^{2n}).$ The fact that   $\mathcal E_{\rm even}^\dagger, $ and $\mathcal E_{\rm odd}^\dagger$ are bounded from  $L^1(\mathbb R, \mathbb C^{2n})$  into $L^1(\mathbb R^+, \mathbb C^{2n})$ and from  $L^\infty(\mathbb R, \mathbb C^{2n})$  into $L^\infty(\mathbb R^+, \mathbb C^{2n}),$ and that $\mathcal R^\dagger$ is bounded  from  $L^1(\mathbb R^+, \mathbb C^{2n})$  into $L^1(\mathbb R, \mathbb C^{2n})$ and from  $L^\infty(\mathbb R^+, \mathbb C^{2n})$  into $L^\infty(\mathbb R, \mathbb C^{2n}),$ follows immediately. Further, 
as
$
  \| P^\dagger_{\pm,\mathcal M_1}\|_{L^1(\mathbb R, M_{2n})}=   \| P_{\pm, \mathcal M_1}\|_{L^1(\mathbb R, M_{2n})},
$  
and since we already proved,
$
P_{\pm, \mathcal M_1} \in L^1(\mathbb R^+, M_{2n}),
$
we obtain,
$
P_{\pm, \mathcal M_1}^\dagger \in L^1(\mathbb R^+, M_{2n}),
$
and then as
 $$
 Q(P_{\pm,\mathcal M_1})^\dagger Y(x)= \int_{-\infty}^\infty\, P_{\pm,\mathcal M_1}^\dagger(y-x)\, Y(y)\,dy,
 $$ 
 it follows that $Q_\pm(P_{\mathcal M_1})^\dagger$ is bounded in $L^1(\mathbb R, \mathbb C^{2n})$ and  in $L^\infty(\mathbb R,\mathbb C^{2n}).$ Finally, it follows from \eqref{new.39.a}
 that  the wave operators  $W_\pm(H,H_1)^\dagger$ extend to bounded operators on $L^1(\mathbb R^+,\mathbb C^{2n})$ and to bounded operators on $L^\infty(\mathbb R^+,\mathbb C^{2n}),$ and by \eqref{new.24} this also holds for the wave operators $W_\pm^\dagger(H,H_1).$ 
 
 \bull
 
 Using the unitary transformation $\mathbf U$ given in \eqref{unitary} we obtain our result on the boundedness of the wave operators on the line, from Theorem~\ref{theo.new.4}. However, since Theorem~\ref{theo.new.4} involves $S(0),$ and $S_\infty,$ we first introduce some concepts from the stationary scattering theory of matrix Schr\"odinger operators on the line that we quote from \cite{akv}.
 Under the assumption that $\mathcal V \in L^1_1(\mathbb R, \mathbb C^n),$ the  Jost solution from the left, $f_l(k,x), x \in \mathbb R,  k \in \overline{\mathbb C^+},$ is the  $ n \times n$ matrix-valued  solution to the Schr\"odinger equation  on the line,
 \beq\label{seline}
 -\frac{d^2}{d x^2} Y(x)+\mathcal V(x) Y(x)= k^2 Y(x), \qquad x \in \mathbb R,
 \ene 
 that satisfies
 \beq\label{new.41}
 f_l(k,x)= e^{ik x} [ I_n+o(1)],  f_l'(k,x)= e^{ik x} [ik I_n+o(1)], x \to \infty.
 \ene
 Further, for $ k \in \mathbb R \setminus \{0\},$  $f_l(k,x)$ fulfills,
 \beq\label{new.42}
 f_l(k,x)= a_l(k) e^{ikx}+ b_l(k) e^{-ikx}+o(1), \qquad x \to -\infty.
 \ene
Similarly,  Jost solution from the right, $f_r(k,x), x \in \mathbb R, k \in \overline{\mathbb C^+},$  is the  $ n \times n$ matrix-valued solution to the Schr\"odinger equation \eqref{seline}, such that,
\beq\label{new.43}
f_r(k,x)=  e^{-ik x} [ I_n+o(1)], f_r'(k,x)=  e^{-ik x} [ - ik I_n+o(1)], \qquad x \to- \infty.
 \ene
 Moreover, for $ k \in \mathbb R \setminus \{0\},$  $f_r(k,x)$ fulfills,
 \beq\label{new.44}
 f_r(k,x)= a_r(k) e^{-ikx}+ b_r(k) \, e^{ikx}+o(1), \qquad x \to \infty.
 \ene
 The transmission coefficient from the left, $T_l(k),$ and the transmission coefficient from the right, $T_r(k,)$ are defined by
 \beq\label{new.45}
 T_l(k):= \frac{1}{a_l(k)},\qquad T_r(k):= \frac{1}{a_r(k)}.
 \ene
The reflection coefficient from the left, $L(k),$ and the reflection coefficient from the right, $R(k),$ are given by,
\beq\label{new.46}
L(k):= \frac{b_l(k)}{a_l(k)}, \qquad R(k):= \frac{b_r(k)}{a_r(k)}.
\ene
The physical solution from the left, $\Psi_l(k,x),$ is defined as,
\beq\label{new.47}
\Psi_{l}(k,x):= T_l(k)\, f_l(k,x).
\ene
Then, $\Psi_{l}(k,x)$ satisfies,
\beq\label{new.48}
\Psi_l(k,x) = \left\{\begin{array}{l}   T(k) e^{ikx}+o(1), x \to \infty, \\   e^{ikx} + e^{-ikx} L(k)+o(1), x \to- \infty.
  \end{array} \right.
  \ene
  The physical solution from the left corresponds to  a scattering process where a particle is incident from the left with unit amplitude, it is reflected with amplitude 
  $L(k)$ and it is transmitted with amplitude $ T_l(k).$
   Similarly, the physical solution from the right, $\Psi_r(k,x),$ is defined as,
  \beq\label{new.49}
  \Psi_r(k,x):= T_r(k) f_r(k,x).
  \ene
  Hence, $\Psi_r(k,x)$satisfies,
  \beq\label{new.50}
  \Psi_r(k,x) = \left\{\begin{array}{l}     e^{-ikx} + e^{ikx} R(k)+o(1), x \to \infty, \\T_r(k) e^{-ikx}+o(1), x \to -\infty.
  \end{array} \right.
  \ene
  The physical solution from the right corresponds to  a scattering process where a particle is incident from the right with unit amplitude, it is reflected with amplitude 
  $R(k)$ and it is transmitted with amplitude $ T_r(k).$
  
The scattering matrix on the line, $S_{\mathbb R}(k),$ is defined as follows,
\beq\label{new.51}
S_{\mathbb R}(k):= \left\{\begin{array}{lc} T_l(k)& R(k)\\ L(k)& T_r(k)\end{array} \right\}.
\ene
  Using our results, we can directly define the physical solutions from the left and from the right from the physical solution $\Psi(k,x), k \in \mathbb R \setminus\{0\},$ given in \eqref{Physicalsolution}, by means of our unitary transformation $\mathbf U$, given in \eqref{unitary}. We proceed as follows.
Let us denote by $\Psi^{(1)}(k,x)$ the $2n\times n$ matrix with the first $n$  columns of $\Psi(k,x),$ and let $\Psi^{(2)}(k,x)$ be the $2n \times n$ matrix with the second $n$ columns of $\Psi(k,x).$ Then, by \eqref{Jostsolution} and \eqref{Physicalsolution},
 \beq\label{new.52}
  \left\{\mathbf U \Psi^{(1)}\right\}_{i j}= \left\{\begin{array}{l}  e^{-ikx} \delta_{i,j}+ e^{ikx} \{S\}_{i, j}(k)+o(1), x \to \infty, 1\leq i,j \leq n,  \\   e^{-ikx} \{S\}_{n+i,j}(k)+o(1), x \to -\infty, 1 \leq i,j \leq n.
  \end{array} \right.
  \ene
Further, by \eqref{new.50} and \eqref{new.52}, we define,
\beq\label{new.53}
\{ \Psi_r(k,x)\}_{i,j}:= \left\{\mathbf U \Psi^{(1)}\right\}_{i j}, \{T_r\}_{i,j}(k):= \{S\}_{n+i,j}(k), R_{i,j}(k):= \{ S\}_{i,j}(k), 1 \leq i,j \leq n.
\ene
 Moreover,
  \beq\label{new.54}
  \left\{\mathbf U \Psi^{(2)}\right\}_{i j}= \left\{\begin{array}{l}  e^{ikx} \{S\}_{i,n+j}(k)+o(1), x \to \infty, 1 \leq i,j \leq n,\\
  e^{ikx} \delta_{i,j}+ e^{-ikx} \{S\}_{n+i, n+j}(k)+o(1), x \to \infty, 1\leq i,j \leq n.  \\  
  \end{array} \right.
  \ene
 Then, by \eqref{new.48} and \eqref{new.54}, we define, 
 \beq\label{new.55}
\{ \Psi_l(k,x)\}_{i,j}:= \left\{\mathbf U \Psi^{(2)}\right\}_{i j}, \{T_l\}_{i,j}(k):= \{S\}_{i,n+j}(k), L_{i,j}(k):= \{ S\}_{n+i,n+j}(k), 1 \leq i,j \leq n.
\ene
Moreover, by \eqref{new.53}, \eqref{new.55}, we can directly define the  scattering matrix on the line from the scattering matrix on the half line as follows.
\beq\label{new.55.aa}
\begin{array}{l}
S_{\mathbb R}(k):= \left\{\begin{array}{lc} T_l(k)& R(k)\\ L(k)& T_r(k)\end{array} \right\}, \,{\rm where, }\, \{T_r\}_{i,j}(k):= \{S\}_{n+i,j}(k), \\ \\R_{i,j}(k):= \{ S\}_{i,j}(k),
\{T_l\}_{i,j}(k):= \{S\}_{i,n+j}(k), L_{i,j}(k):= \{ S\}_{n+i,n+j}(k), 1 \leq i,j \leq n.
\end{array}
\ene
Note that, by \eqref{new.55.aa}
\beq\label{new.56}
S(0)= S_{\infty}= \left\{ \begin{array}{lc} 0_n& I_n\\ I_n& 0_n\end{array} \right\} \Longleftrightarrow S_{\mathbb R}(0)= S_{\mathbb R, \infty}= \left\{ \begin{array}{lc} I_n& 0_n\\ 0_n& I_n\end{array} \right\},
\ene 
where we denote,
\beq\label{new.57}
S_{\mathbb R, \infty}= \lim_{|k|\to \infty} S_{\mathbb R}(k).
\ene
%

\noindent{\it Proof of Theorem~\ref{theo.new.6}: } By \eqref{hwhole} and \eqref{w.28}
\beq\label{new.59}
W_\pm(H_{\mathbb R}, H_{0,\mathbb R})= \mathbf U W_\pm(H,H_1) \mathbf U^\dagger.
\ene
Then, the theorem follows from Theorem~\ref{theo.new.4} and \eqref{new.56}.

\bull
 \appendix
 \sss
 \renewcommand{\theequation}{\thesection.\arabic{equation}}

\newtheorem{theorem2}{THEOREM}[section]
\renewcommand{\thetheorem}{\arabic{section}.\arabic{theorem}}

\newtheorem{prop2}[theorem2]{PROPOSITION}
\newtheorem{lemma2}[theorem2]{LEMMA}
 \section{The scattering matrix for potentials in $L^1_{\gamma}(\mathbb R^+,M_n),  2 \leq \gamma \leq 3$}
 In this appendix we always assume that  $ V\in  L^1_\gamma(\mathbb R^+,M_n),   2 \leq \gamma \leq 3.$  
  By the definition of $S(k)$ in \eqref{Scatteringmatrix} and by Proposition 3.2.4 in page 65, Theorem 3.81 in page 114 and Theorem 3.9.15 in pages 189-190 of \cite{WederBook} $S(k)$ is  differentiable for $k \in \mathbb R,$  with continuous derivative for $ k \in \mathbb R\setminus\{0\},$ provided that $V \in L^1_2(\mathbb R^+).$ We denote by   $\dot{S}(k)$  the derivative of $S(k).$
    Moreover, by Theorem 3.10.6 in page 196-197 of \cite{WederBook},
  \beq\label{aa.1}
  S(k)-S_\infty=O\left( \frac{1}{|k|}\right), \qquad |k| \to \infty.
  \ene
  We now consider  the high-energy behavior of the derivative of $S(k).$ 
  \begin{prop2}
  suppose that $\eqref{PotentialHermitian}$ is satisfied, that $ V\in L^1_2(\mathbb R^+),$ and that the constant  matrices $A,B$ satisfy \eqref{wcon1}, and \eqref{wcon2}.
 Then,
  \beq\label{abb.1}
 \dot{S}(k)= O\left( \frac{1}{|k|}\right), \qquad |k| \to \infty.
\ene
\end{prop2} 

\noindent{\it Proof:}
  As in \cite{WederBook} we denote, $m(k,x):= e^{-ikx} f(k,x).$
  By Proposition 3.9.1 in pages 155-156 of \cite{WederBook} $\dot{m}(k,x),$ and $\dot{m}'(k,x)$ exist for $ x \in [0,\infty)$ and $ k \in \overline{\mathbb C^+},$ they are analytic in $ k \in \mathbb C^+$ and continuous in $k \in \overline{\mathbb C^+}$ for each $x \in [0,\infty),$ and they are continuous in $x \in[0,\infty)$ for each $ k\in \overline{\mathbb C^+}.$ Moreover, by equation (3.2.30) in page 56 of \cite{WederBook},
  \beq\label{aa.2}
  m(k,0)= I+ O\left( \frac{1}{|k|}\right),  \qquad  k \to \infty \, \text{\rm in} \, \overline{\mathbb C^+}. 
  \ene
  Further, by equations  (3.9.3)  and (3.94) in page 156, and equations (3.9.15 ) in page 157  and (3.9.17)  in page 158 of \cite{WederBook}, and \eqref{aa.2}
  \beq\label{aa.3}
  \dot{m}(k,0)=  O\left( \frac{1}{|k|}\right),   \dot{m}'(k,0)= O( 1),  \qquad   k \to \infty \, \text{\rm in} \, \overline{\mathbb C^+}.  
  \ene
  By \eqref{Jostmatrix}
 \beq\label{aa.4}
  \dot{J}(k)= -\dot{m}(-k,0)^\dagger B- i m(-k,0)^\dagger A+ i k \dot{m}(-k,0)^\dagger A+ \dot{m}'(-k,0)^\dagger A , \qquad k \in \mathbb R.
  \ene
  Then, by \eqref{aa.2}-\eqref{aa.4}
  \beq\label{aa.5}
  \dot{J}(k)= O(1) A+ O\left( \frac{1}{|k|}\right),  \qquad  |k| \to \infty \, \text{\rm in} \, \mathbb R.   
  \ene
  Further, by \eqref{Scatteringmatrix}
  \begin{align}\nonumber
  \dot{S}(k)&=  \left( \dot{J}(-k)J_0(-k)^{-1}\right)   \left(J_0(-k) J_0(k)^{-1}\right) \left(J_0(k)  J(k)^{-1}\right)  + \\\label{aa.6}
  &(J(-k) J_0(-k)^{-1})\, (J_0(-k) J_0(k)^{-1} ) (J_0(k)  J(k)^{-1})  \left(\dot{J}(k) J_0(k)^{-1}\right)  \left( J_0(k) J(k)^{-1} \right).  
  \end{align}
 By \eqref{aa.5} and equations (3.7.11) and  (3.7.12) in page 113 of \cite{WederBook},
 \beq\label{aa.7}
 \dot{J}(-k)J_0(-k)^{-1}= O\left( \frac{1}{|k|}\right),  \qquad   |k| \to \infty \, \text{\rm in} \, \mathbb R. 
 \ene
 Moreover, by equations (3.6.3) in page 110, and (3.7.3), (3.7.4) in page 113 of \cite{WederBook}, 
 \beq\label{aa.8}
J_0(-k) J_0(k)^{-1} = O(1), \qquad  k \to \infty \, \text{\rm in} \,\mathbb C.
\ene
Further, by  equations (3.10.17) and (3.10.18) in page 194 of \cite{WederBook},
\beq\label{aa.9}
J(k) J_0(k)^{-1}= I+ O\left( \frac{1}{|k|}\right), J_0(k) J(k)^{-1}=I+O\left( \frac{1}{|k|}\right), \qquad  k \to \infty \, \text{\rm in} \, \overline{\mathbb C^+}. 
\ene
Finally, by \eqref{aa.6}-\eqref{aa.9} we obtain,
\beq\label{aa.10}
\dot{S}(k)= O\left( \frac{1}{|k|}\right), \qquad  |k| \to \infty. 
\ene

\bull

We now study the low-energy behavior of  $\dot{S}(k).$

\begin{prop2}\label{prople}
  suppose that $\eqref{PotentialHermitian}$  holds  and that the constant  matrices $A,B$ satisfy \eqref{wcon1}, and \eqref{wcon2}.Then,
  \begin{itemize}
  \item[a)]
  In the {\it generic case} where $J(0)$ is invertible, if $ V\in L^1_2(\mathbb R^+),$
  \beq\label{aabbb}
  \dot{S}(k)= O\left( 1\right), \qquad |k| \to 0.
  \ene
  \item[b)]
  In the {\it exceptional case} where $J(0)$ is not invertible, if  $ V\in L^1_\gamma(\mathbb R^+), 2 \leq \gamma \leq 3,$ 
  \beq\label{abb.2}
\dot{S}(k)= O\left( |k|^{\gamma -3}\right), \qquad |k| \to 0.
\ene
\end{itemize}
\end{prop2} 
\noindent{\it Proof:}   As in \cite{WederBook} we denote, $m(k,x):= e^{-ikx} f(k,x).$ Note that $m(0,x)= f(0,x).$
By equations (3.2.13), (3.2.14), (3.2.15) and (3.2.16) in page 54 of \cite{WederBook}, and since,
\beq\label{abb.0}
| e^{z}-1|\leq C \frac{|z|}{1+|z|}, \qquad z \in \mathbf C,
\ene
we have
\beq\label{abb.00}
| m(k,x)| \leq C, \qquad k \in \overline{\mathbb C^+}, x \in \mathbb R^+,
\ene
provided that $ V\in L^1_1(\mathbb R^+).$
By equation (3.9.15) in page 157 of \cite{WederBook}, 
\beq\label{abb.3}
|\dot{m}(k,x)| \leq C, \qquad k \in \overline {\mathbb C^+}, x \in \mathbb R^+.
\ene

By equation (3.9.17) in page 158 of \cite{WederBook},
\beq\label{dddd}
\left| \dot{m}'(k,x)\right| \leq C, \qquad k \in \overline{\mathbb C^+}, x \in \mathbb R^+.
\ene

Further, by \eqref{abb.00} and  \eqref{abb.3}
\beq \label{abb.4}
\left| m(k,x)- m(0,x)\right| \leq C \hbox{\rm min} [|k|, 1], \qquad  k \in \overline{\mathbb C^+}, x \in \mathbb R^+.
\ene
Moreover, by equation (3.9.239) in page  190 of \cite{WederBook},
\beq\label{abb.17}
S(k)= S(0)+ k \dot{S}(0)+o\left( |k|\right), \qquad k \to 0.
\ene

By \eqref{Scatteringmatrix}, 
\beq\label{abb.18}
\dot{S}(k)= \dot{J}(-k)\, J(k)^{-1}+ J(-k)\, J(k)^{-1} \, \dot{J}(k) \,J(k)^{-1}=\dot{J}(-k)\, J(k)^{-1}- S(k)\, \dot{J}(k) \,J(k)^{-1}.
\ene

If $J(0)$ is invertible, \eqref{aabbb} follows from Proposition~\ref{Jostnozero}, \eqref{aa.4}, \eqref{abb.00}, \eqref{abb.3}, \eqref{dddd}, and the first equality in \eqref{abb.18}. This proves item a). Let us prove b).
By equation (3.9.3) in page 156 of \cite{WederBook}
\beq\label{abb.5}
\dot{m}(k,x)= \dot{m}_0(k,x)+ \frac{1}{2ik}\, \int_{x}^\infty\, dy\,\left[ e^{2i k(y-x)}-1 \right]\, V(y)\, \dot{m}(k,y),
\ene
where,
\beq\label{abb.6}
\dot{m}_0(k,x):=\frac{1}{2i k^2}\, \int_x^\infty\, dy\, \left[ e^{-2i k(y-x)} -1+2ik (y-x)\right]  e^{2ik(y-x)}\, V(y) \, m(k,y).
\ene
Further, taking the limit as $ k \to 0$ in \eqref{abb.5} and \eqref{abb.6}, and using \eqref{abb.00}  and \eqref{abb.3} we obtain
\beq\label{abb.7}
\dot{m}(0,x)= \dot{m}_0(0,x)+  \int_{x}^\infty\, dy\,(y-x)\, V(y)\, \dot{m}(0,y), 
\ene
with,
\beq\label{abb.8}
\dot{m}_0(0,x):=  i \int_x^\infty\, dy\, (y-x)^2\, e^{2ik(y-x)}\, V(y) \, m(0,y).
\ene
Note that
\beq\label{abb.9}
|e^z-1-z| \leq C \frac{|z|^2}{1+|z|}, \qquad z \in \mathbf C,
\ene
and
\beq\label{abb.10}
| e^z-1-z- \frac{z^2}{2}| \leq  C \frac{|z|^3}{1+|z|}, \qquad z \in \mathbf C.
\ene
It follows from \eqref{abb.0}, \eqref{abb.4}, \eqref{abb.6}, \eqref{abb.8},\eqref{abb.9} and \eqref{abb.10} that,
\beq\label{abb.11}
|\dot{m}_0(k,x)-\dot{m}_0(0,x)| \leq C \,  \hbox{\rm min} [ |k|^{\gamma-2}, 1], \qquad k \in \overline{\mathbb C^+}, x \in \mathbb R^+.
\ene
By equations (3.9.6) and (3.9.7) in page 156 of \cite{WederBook},
\beq\label{abb.11b}
\dot{m}(k,x)= \sum_{j=0}^\infty\, \dot{m}_j(k,x),
\ene
where
\beq\label{abb.11c}
\dot{m}_j(k,x):= \frac{1}{2ik}\, \int_x^\infty\, dy\, \left[e^{2ik(y-x)}-1\right]\, V(y)\, \dot{m}_{j-1}(k,x), \qquad j \geq 1,
 \ene
 and the series in \eqref{abb.11b} is uniformly convergent. Taking the limit as $ k \to 0$ in \eqref{abb.11b} and \eqref{abb.11c} we get,
 \beq\label{abb.11d}
\dot{m}(0,x)= \sum_{j=0}^\infty\, \dot{m}_{j}(0,x),
\ene
where
\beq\label{abb.11e}
\dot{m}_j(0,x):=  \int_x^\infty\, dy\, (y-x)\, V(y)\, \dot{m}_{j-1}(0,x), \qquad j \geq 1.
 \ene
By \eqref{abb.11b} and \eqref{abb.11d},
\beq\label{abb.11ff}
\dot{m}(k,x)-\dot{m}(0,x)=\dot{m}_0(k,x)-\dot{m}_0(0,x)+ \sum_{j=1}^\infty\,(\dot{m}_j(k,x)- \dot{m}_j(0,x)),
\ene 
and by \eqref{abb.11c} and \eqref{abb.11e}, for $ j \geq 1,$
\beq\label{abb.11g}
\dot{m}_j(k,x)- \dot{m}_j(0,x)=
\int_x^\infty\, dy\, V(y)\, \left( \frac{1}{2ik}\, \left[ e^{2i k(y-x)} -1 \right]\, \dot{m}_{j-1}(k,x) - (y-x)\, \dot{m}_{j-1}(0,x)\right).
\ene
By equation (3.9.11) in page 156 of \cite{WederBook}
\beq\label{abb.11h.a}
|\dot{m}_0(k,x)| \leq C, \qquad k \in \overline{\mathbb C^+}, x \in \mathbb R^+.
\ene
Further, by equation (3.9.12) in page 157 of \cite{WederBook},
\beq\label{abb.11h}
|\dot{m}_j(k,x)| \leq \int_x^\infty\, dy\, y\, |V(y)|\, |\dot{m}_{j-1}(k,y)|, \qquad j \geq 1, \qquad k \in \overline{\mathbb C^+}, x \in \mathbb R^+.
\ene
Then, by \eqref{abb.9}, and  \eqref{abb.11g}, for $ j \geq 1,$
\beq\label{abb.11i}\begin{array}{l}
|\dot{m}_j(k,x)- \dot{m}_j(0,x)| \leq C\,\int_x^\infty\, dy\,(1+ y)^2\, |V(y)|\,  \left[ |k|\, |\dot{m}_{j-1}(k,y)|+|\dot{m}_{j-1}(k,y)- \dot{m}_{j-1}(0,y)|\right].
\end{array}
\ene
Without  loss of generality we can take the constant $C$ in \eqref{abb.11}, \eqref{abb.11h.a}, and \eqref{abb.11i} bigger or equal than one. Then, using \eqref{abb.11}, \eqref{abb.11h.a},\eqref{abb.11h}, and \eqref{abb.11i}, we prove by mathematical induction that for $ j \geq 0$
\beq\label{abb.11j}
|\dot{m}_j(k,x)- \dot{m}_j(0,x)| \leq \hbox{\rm min} [ |k|^{\gamma-2}, 1]\,
  (j+1)\, C^{j+1}\, \frac{1}{j!}\, \left[ \int_x^\infty\, dy\, (1+y)^2 |V(y)| \right]^j, \qquad  k \in \overline{\mathbb C^+}, x \in \mathbb R^+.
\ene
Then, by \eqref{abb.11ff} and \eqref{abb.11j}
\beq\label{abb.11f}\begin{array}{l}
|\dot{m}(k,x)- \dot{m}(0,x)| \leq C\displaystyle  \hbox{\rm min} [ |k|^{\gamma-2}, 1]\,e^{C \int_x^\infty\, dy\,(1+y)^2\, |V(y)| }\left[1+\right. \\ \\\left.
C   \,  \int_x^\infty\, dy\, (1+y)^2\, |V(y)|\right] \leq C_1\,\hbox{\rm min} [ |k|^{\gamma-2}, 1]\qquad k \in \overline{\mathbb C^+}, x \in \mathbb R^+,
\end{array}
\ene
for a constant $C_1.$

Furthermore, taking the derivative with respect to $k$ in both sides of equation (3.2.7) in page 53 of \cite{WederBook} we get
\beq\label{abb.12}
\dot{m}'(k,x)= -2i  \int_x^\infty\, dy \, (y-x) \,e^{2ik(y-x)}\, V(y)\, m(k,y) -\int_x^\infty\, dy\,e^{2i k(y-x)} \,V(y)\, \dot{m}(k,y).
\ene
Taking the limit as $ k \to 0,$ in \eqref{abb.12} and using \eqref{abb.00} and \eqref{abb.3} we obtain,
\beq\label{abb.13}
\dot{m}'(0,x)= -2i  \int_x^\infty\, dy \, (y-x) \,V(y)\, m(0,y) -\int_x^\infty\, dy\,  V(y)  \dot{m}(0,y).
\ene
By \eqref{abb.0}, \eqref{abb.00}, \eqref{abb.3}, \eqref{abb.4},\eqref{abb.11f}, \eqref{abb.12}, and \eqref{abb.13}, it follows that,
\beq\label{abb.14}
|\dot{m}'(k,x)-\dot{m}'(0,x)| \leq C \,  \hbox{\rm min} [ |k|^{\gamma-2}, 1], \qquad  k \in \overline{\mathbb C^+}, x \in \mathbb R^+. 
\ene
Furthermore, by \eqref{aa.4}, \eqref{abb.3}, \eqref{abb.4}, \eqref{abb.11f},  and \eqref{abb.14},
\beq\label{abb.15}
\dot{J}(k)-\dot{J}(0)= O\left( |k|^{\gamma-2} \right), \qquad |k| \to 0\,  \text{\rm in }\, \mathbb R.
\ene
By equation (3.9.237) in page 189 of \cite{WederBook},
\beq\label{abb.16}
J(k)^{-1}= \frac{1}{k} \, \mathcal M+ \mathcal E_1+o(1),  \qquad k\to 0 \, \text{\rm in}\, \overline{\mathbb C^+} ,
\ene
where $\mathcal M$ and $\mathcal E_1$ are constant matrices. 
Then, by \eqref{abb.17}, the second equality in \eqref{abb.18}, \eqref{abb.15}, and   \eqref{abb.16},
\beq\label{aab.19}
\dot{S}(k)=  \frac{1}{k}\, \mathcal N+ O\left( |k|^{\gamma-3} \right ), \qquad k \to 0,
\ene
where,
\beq\label{aab.20}
\mathcal N:= \dot{J}(0) \mathcal M -S(0) \dot{J}(0)\, \mathcal M.
\ene
In the case $ \gamma=2,$ \eqref{aab.19} gives us \eqref{abb.2}. When, $ 2 < \gamma \leq 3,$ from \eqref{aab.19} we obtain, for $\varepsilon >0,$
\beq\label{aab.21}
S(\varepsilon)= S(1)-\int_\varepsilon^1\, dk\, \dot{S}(k)= S(1)+\ln \varepsilon \, \mathcal N+ O(1), \varepsilon \downarrow0.
\ene
However, \eqref{aab.21} is compatible with \eqref{abb.17}  only if  $\mathcal N=0.$ Hence, by \eqref{aab.19},
\beq\label{aab.22}
\dot{S}(k)=O\left(|k|^{\gamma-3}\right), \qquad k \to 0.
\ene
This concludes the proof of \eqref{abb.2}.

\bull

 The results above give us the following proposition
 \begin{prop2}\label{prop.h1}
 Suppose that $V$ fulfills \eqref{PotentialHermitian} and that the constant  matrices $A,B$ satisfy \eqref{wcon1}, and \eqref{wcon2}.
 Then,  $S(k)-S_\infty \in \mathbf H^{(1)}(\mathbb R, M_n),$ provided that in the {\it generic case}, where $J(0)$ is invertible,$ V\in L^1_2(\mathbb R^+),$ and in the {\it exceptional case}, where $J(0)$ is not invertible,  $ V \in L^1_\gamma(\mathbb R^+), \gamma  > \frac{5}{2}.$ 
 \end{prop2}
 \noindent{\it Proof:}  Since $S(k)$ is differentiable  for $ k \in \mathbb R,$ with continuous derivative for $ k \in \mathbb R\setminus\{0 \}, $ and it satisfies \eqref{aa.1} \eqref{abb.1}, \eqref{aabbb}, and \eqref{abb.2},  it follows that it belongs to 
 $\mathbf H^{(1)}(\mathbb R^+, M_n).$
 
  \qed

  \noindent{\bf Acknowledgement}
  
\noindent This paper was partially written while I was visiting the Institut  de\linebreak  Math\'ematique d'Orsay, Universit\'e  Paris-Sud. I thank Christian G\'erard for his kind hospitality

\end{document}